\documentclass[12pt,reqno,a4paper]{article}

\usepackage{amsmath,amsthm,amstext,amscd,amssymb,euscript,mathrsfs}
\usepackage{calrsfs}
\usepackage{epsfig}
\usepackage{ dsfont }
\usepackage{tikz}
\usetikzlibrary{decorations.pathreplacing,calligraphy}
\usepackage[blocks]{authblk}
\usepackage{hyperref}
\usepackage{subfig}
\usepackage{float}
\usepackage{footmisc}

\usepackage{geometry}
\newcommand{\Z}{\mathbb Z}
\newcommand{\R}{\mathbb R}
\newcommand{\N}{\mathbb N}

\newcommand{\E}{\mathbb E}

\renewcommand{\Pr}{\mathbb P}

\renewcommand{\phi}{\varphi}

\newcommand{\om}{\ensuremath{\omega}}

\newcommand{\pee}{\ensuremath{\mathbb{P}}}

\def\1{{\mathchoice {\rm 1\mskip-4mu l} {\rm 1\mskip-4mu l}
{\rm 1\mskip-4.5mu l} {\rm 1\mskip-5mu l}}}

\newtheorem{theorem}{{\small T}{\scriptsize HEOREM}}[section]
\newtheorem{corollary}{{\bf{\small C}{\scriptsize OROLLARY}}}[section]
\newtheorem{proposition}{{\bf{\small P}{\scriptsize ROPOSITION}}}[section]
\newtheorem{lemma}{{\bf{\small L}{\scriptsize EMMA}}}[section]
\newtheorem{remark}{{\bf{\small R}{\scriptsize EMARK}}}[section]
\newtheorem{definition}{{\bf{\small D}{\scriptsize EFINITION}}}[section]
\newtheorem{example}{{\bf{\small E}{\scriptsize XAMPLE}}}[section]

\renewenvironment{proof}[1]
{\noindent{{\bf{\small{ P}{\scriptsize ROOF}}}.}\hspace{0.1cm} #1} {$\;\qed$\newline}

\newcommand{\beq}{\begin{eqnarray}}
\newcommand{\eeq}{\end{eqnarray}}

\newcommand{\ba}{\begin{align*}}
\newcommand{\ea}{\end{align*}}

\newcommand{\be}{\begin{equation}}
\newcommand{\ee}{\end{equation}}

\newcommand{\bl}{\begin{lemma}}
\newcommand{\el}{\end{lemma}}

\newcommand{\br}{\begin{remark}}
\newcommand{\er}{\end{remark}}

\newcommand{\bt}{\begin{theorem}}
\newcommand{\et}{\end{theorem}}

\newcommand{\bd}{\begin{definition}}
\newcommand{\ed}{\end{definition}}

\newcommand{\bp}{\begin{proposition}}
\newcommand{\ep}{\end{proposition}}

\newcommand{\bc}{\begin{corollary}}
\newcommand{\ec}{\end{corollary}}

\newcommand{\bpr}{\begin{proof}}
\newcommand{\epr}{\end{proof}}

\newcommand{\bi}{\begin{itemize}}
\newcommand{\ei}{\end{itemize}}

\newcommand{\ben}{\begin{enumerate}}
\newcommand{\een}{\end{enumerate}}

\newcommand{\caE}{{\mathrsfs E}}

\newcommand{\caP}{{\mathcal P}}

\makeatletter
\let\@fnsymbol\@roman
\makeatother

\begin{document}

\title{Random walks in a field of soft traps and criticality for the dissipative Abelian Sandpile Model}

\author[a,1]{Frank Redig}
\author[b,2]{Ellen Saada}
\author[a,3]{Berend van Tol}

\affil[a]{Institute of Applied Mathematics, Delft University of Technology, The Netherlands}
\affil[b]{Laboratoire MAP5, Université Paris Cité, Paris, France}

\footnotetext[1]{Email: \texttt{F.H.J.Redig@tudelft.nl}}
\footnotetext[2]{Email: \texttt{ellen.saada@mi.parisdescartes.fr}}
\footnotetext[2]{Email: \texttt{B.T.vanTol@tudelft.nl}}

\date{\today}

\maketitle
\begin{abstract}
Motivated by the dissipative abelian sandpile model, we  analyze the trajectories of a one-dimensional random walk in a landscape of soft traps.  These traps, placed at increasing distances from each other,  correspond to dissipative sites in the associated dissipative abelian sandpile model.
We identify a critical growth rate of the sizes of intervals between successive traps where there is a transition between finiteness and non-finiteness of the expected survival time of the random walk. This corresponds to a transition between non-criticality and criticality of the associated dissipative abelian sandpile model. Therefore, in this setting, we thus identify precisely how much dissipation can be added to the original abelian sandpile model in order to disrupt its criticality.
\end{abstract}

\section{Introduction}\label{sec:intro}
The abelian sandpile \cite{btw}, \cite{dhar} is a well-known model of self-organized criticality.
We briefly summarize the original model; for more details about the definition of the model and its properties, we refer to \cite{dhar}.
On a finite lattice, a height configuration is such that on each site there is a maximal allowed height, that is, a maximal
number of allowed grains. At each time step,
a sand grain is uniformly added on a site, say $x$; if the height on $x$ becomes larger than allowed, $x$ does a \textit{toppling}, 
sending one sand grain to each of its neighboring sites, plus (at least) one grain outside of the system if $x$ is a boundary site (in such a case $x$ is said to be \textit{dissipative}); 
then, if on a neighboring site $y$ of $x$ there are now too many grains, $y$ topples, and so on
(this is an \textit{avalanche}), until the system is stabilized, that is, all the sites have allowed heights. Stabilization is always possible since the lattice is finite and it is assumed that from every site there is a  nearest neighbor
path towards a dissipative site. A step of the Markovian dynamics consists in starting
from a stable configuration, and
in one time step choosing a uniform random site $x$ and passing to the stabilization of the configuration where one grain is added to $x$.

In the original model, only boundary sites are dissipative, i.e., there is a loss of mass upon toppling those sites. 
The model exhibits self-organized criticality, which means that in the infinite-volume limit of the stationary state, the avalanche size has a power-law decay and has infinite expectation (comparable to a critical percolation cluster). Another manifestation of self-organized criticality is that in the stationary state there is power-law decay of covariances of the height variables.

The criticality of the model is strongly related to the fact that only boundary sites are dissipative, i.e., in  the bulk toppling conserves the total height.
Indeed, in \cite{katori} a dissipative version of the abelian sandpile was introduced and studied, and the authors showed that introducing bulk dissipation leads to loss of self-organized criticality as manifested e.g. by the fact that the tail of the avalanche size distribution decays exponentially, and avalanches have finite first moment.
In \cite{maes2004infinite}, the infinite-volume limit of this dissipative model was studied. In \cite{jarai2015approaching} the model with continuous heights is studied (i.e., where the height value at each site is a non-negative real number). One can then consider arbitrary small dissipation, and it was shown that the limit of zero dissipation coincides with the original critical model.
In \cite{redig2018non} was raised the question of how much dissipation has to be added to the original abelian sandpile model in order to lose its self-organized criticality. The loss of criticality was defined in \cite{redig2018non} as having finite expected avalanche sizes in the infinite-volume limit. Therefore,  the question raised in \cite{redig2018non} consists of characterizing where is the transition between finite and infinite expected avalanche size.

In the simplest version of the model, on a one-dimensional chain, adding a finite number of dissipative sites does not break the criticality.
Even if after adding a finite number of dissipative sites avalanches become almost surely finite, their expected size is infinite, because they are proportional to the return time of a one-dimensional simple random walk to its starting point, which is finite almost surely (by recurrence) but has an infinite first moment.
However, adding a positive density of dissipative sites does break the criticality.

In \cite{redig2018non}  the non-criticality of the model with dissipative sites was shown to be equivalent with the finiteness of the expected survival time of an associated trapped random walk, where dissipative sites are exactly the (soft) traps where the walker is killed with a probability proportional to the amount of dissipation.

This setting of the dissipative abelian sandpile model is the main motivation for our paper.
We consider a random walk on the one-dimensional integer lattice, in a landscape of (soft) traps, where the distances between successive traps are  (rapidly) increasing. In particular, the density of traps is zero. We are then interested in identifying the ``critical growth rate'' of the distances between successive traps, where, as a function of a parameter, there is a transition between the expected survival time being finite or infinite. The corresponding dissipative one-dimensional abelian sandpile model is then the one where on a non-dissipative site the stable heights are $0,1$ and upon toppling two grains are distributed, one to each neighbor, and a dissipative site can have stable heights
$0,1,2$ and upon toppling it, two grains are distributed, one to each neighbor, and one grain is lost.
In the language of the corresponding trapped random walk model, on a non-dissipative site, the walk moves left or right with equal probability, and on a dissipative site, it moves left or right with probability $1/3$ and is killed with probability $1/3$.
Denoting $x_i$ the location of the $i$-th trap (we will assume $x_0=0$, i.e., there is a trap at the origin), we are then interested in the critical growth rate of the interval sizes  $|I_i|= x_i-x_{i-1}$.

Our main result (Theorem \ref{thm:main_result}) is that if the growth rate is described by the recursion $|I_{i+1}|= c|I_i|^2$  for some constant $c>0$,  then as a function of $c> 1/(|I_1|)^{-1}$, there is a transition between the expected survival time finite (for $c<1$) or infinite (for $c>1$). This, as explained above, corresponds to a transition between non-criticality (for $c<1$) or criticality (for $c>1$) of the associated dissipative abelian sandpile model.  We thus identify the border of the  ``amount of dissipation'' which is needed to break criticality. 
The growth rate prescribed by the recursion $|I_{i+1}|= c|I_i|^2$ implies that the interval sizes between successive traps increase super-exponentially, which implies in particular that the density of traps is zero. Our proofs are based on a thorough study of the trajectories of the random walk, and we believe it will help us to derive more results on the dissipative sandpile model in the future.\\

Finally, let us mention a subsequent paper \cite{rez}. This paper studies the phase transition between criticality and non-criticality in the dissipative sandpile model in dimension one in the same setting via a different approach. \\

The rest of our paper is organized as follows.
In Section \ref{sec:Random_walk_in_a_landscape_of_soft_traps} we introduce the random walk in a landscape of soft traps, and basic associated quantities, the hitting and inter-arrival times, i.e., times between hitting two successive traps. In Section \ref{sec:main_results} we state our main result (Theorem \ref{thm:main_result}), i.e., identifying,  through the critical
recursion $|I_{i+1}|= c|I_i|^2$, a phase transition according to the value of the constant $c$.  Section \ref{sec:proofofmain} is devoted to the proof of Theorem \ref{thm:main_result}. Here an essential step is to consider the embedded random walk moving between successive trap locations, and to estimate the effect of changing embedded random walk increments on the expected inter-arrival times. This step is the content of Lemma \ref{lemma:contributions}, which is combined with an explicit formula for the expected inter-arrival times when all the embedded random walk increments are to the right.
Finally, to arrive at critical value $c=1$, we prove that the finiteness of the expected survival time is 
a ``tail property'' of the trap configuration, i.e., it does not depend of the locations of the first $n$ traps. 
In Section \ref{sec:Monotonicity} we discuss the (non)-monotonicity of the expected survival time as a function of the sizes of intervals between successive traps.

\section{Random walk in a landscape of soft traps} \label{sec:Random_walk_in_a_landscape_of_soft_traps}
 We consider a simple random walk $\{S_n,n\in\N\}$
on the set of integers $\N=\{0,1,\cdots\} $, reflected at the origin.
The random walk starts at the origin, i.e., $S_0=0$, and the transition probabilities of the walk are defined as follows.
\begin{eqnarray}\label{eq:transprob}
\mathbb{P} (S_{n+1}=y|S_n=x)=
\begin{cases}
1\ \text{if}\ x=0, y=1,\\
1/2\ \text{if}\ x\geq 1, |y-x|=1.
\end{cases}
\end{eqnarray}
This walk will be moving through a landscape of traps.
The trap configuration is given by $\om: \N \rightarrow \{0,1\}$, where $\om(x) = 1$ if there is a trap at $x \in \N$ and $\om(x) = 0$ otherwise. We assume that there is always a trap at zero, i.e. $\om(0) = 1$. The random walk starts at the origin and moves on the integers until it hits a site with a trap. Upon hitting the trap the walker \emph{dies} with probability $1/3$. If the walker survives it resumes according to the transition probabilities \eqref{eq:transprob}.
To formalize this we introduce the extra absorbing state $\ast$. Then the trapped random walk is the process $\{S^{\om}_n,n\in\N\}$ on $\N\cup \{\ast\}$ with transition probabilities as follows.
\begin{eqnarray}\label{eq:transprobom}
\mathbb{P} (S^\om_{n+1}=y|S^\om_n=x)=
\begin{cases}
1/2\ \text{if}\ x\geq 1, |y-x|=1, \text{and}\ \om(x)=0,\\
1/3\ \text{if}\ x\in\N, y=\ast, \ \text{and}\ \om(x)=1,\\
2/3\ \text{if}\ x=0, y=1,
\\
1/3\ \text{if}\ x\geq 1, |y-x|=1, \text{and}\ \om(x)=1,\\
1\ \text{if}\ x=\ast, y=\ast .
\end{cases}
\end{eqnarray}
Usually, to alleviate notation,  we will omit the index $\omega$ from the trapped random walk whenever the context is clear.
We denote by $\pee$ the path space measure of the process $\{S^\om_k, k\geq 0\}$ and by $\E$ the corresponding expectation. 
We define the survival time as
\be\label{eq:survivaltime}
\tau = \inf\{k \geq 0: S_k = \ast\}
\ee
and we say that the trap landscape $\om$ is \emph{critical} if $\E(\tau) = \infty$, \emph{non-critical} if $\E(\tau) < \infty$.

 We further define: 
\begin{itemize}
    \item[(i)] The trap locations: $x_0=0$ and, for $i\geq 0$, $x_{i+1}$ is the first location $x$ to the right of $x_i$ where $\omega(x)=1$. In this way the locations of the traps are $x_0=0<x_1<x_2\ldots$. For trap location $x_i$, we call $i$ the corresponding trap number.
    We further denote the set of traps by
    \be\label{eq:D} 
    D= \{x_0, x_1, x_2, \ldots\}= \{x: \omega(x)=1\},\ee 
    and the number of traps visited before reaching $\ast$ as
    
    \be\label{eq:N}
    N= \sum_{i=0}^\tau \mathbb{I}(S^\om_j\in D).
    \ee
  In \eqref{eq:N} and throughout this paper we use $\mathbb{I}$ 
to denote the indicator function.

\item[(ii)] The intervals between traps:
\be\label{eq:Ii}I_i := [x_{i-1},x_i) \cap \N,\qquad\hbox{for } i\geq 1.\ee

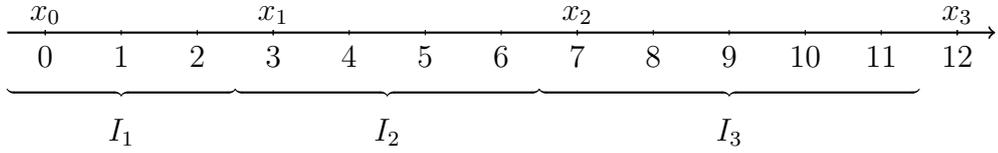
\begin{center}
    \begin{tikzpicture}
\centering
\draw[thick,->] (-0.5,0) -- (12.5,0) ;
\foreach \x in {0,1,2,3,4,5,6,7,8,9,10,11,12}
   \draw (\x cm,1pt) -- (\x cm,-1pt) node[anchor=north] {$\x$};
   
\draw (0 cm,1pt) -- (0 cm,-1pt) node[anchor=south] {$x_0$};
\draw (3 cm,1pt) -- (3 cm,-1pt) node[anchor=south] {$x_1$};
\draw (7 cm,1pt) -- (7 cm,-1pt) node[anchor=south] {$x_2$};
\draw (12 cm,1pt) -- (12 cm,-1pt) node[anchor=south] {$x_3$};

\draw [thick, decorate, decoration = {calligraphic brace,mirror}] (-0.5,-0.75) --  (2.5,-0.75) node[pos=0.5,below = 7pt,black]{$I_1$};

\draw [thick, decorate, decoration = {calligraphic brace,mirror}] (2.5,-0.75) --  (6.5,-0.75) node[pos=0.5,below = 7pt,black]{$I_2$};
\draw [thick, decorate, decoration = {calligraphic brace,mirror}] (6.5,-0.75) --  (11.5,-0.75) node[pos=0.5,below = 7pt,black]{$I_3$};
\end{tikzpicture}
\end{center}

\item[(iii)] The successive hitting times of trap locations 
\be\label{eq:hiti}
\mathcal{T}_i := \inf\left\{k \geq 1: \sum_{j=1}^k \mathbb{I}(S^\om_j \in D) = i\right\},
\ee
for $i \geq 1$. Additionally, we put
$\mathcal{T}_0 := 0$. We then define the associated inter-arrival times 
\be\label{eq:Yi}Y_i:= \mathcal{T}_i - \mathcal{T}_{i-1}\quad\hbox{for } 1 \leq i \leq N,\qquad\hbox{with }\quad Y_i = 0\quad\hbox{for } i > N.\ee
\end{itemize}
We finally define the \emph{embedded} random walk $(\xi_n)_{n\geq0}$ as 
\be\label{eq:embedded}
\xi_i := 
\begin{cases}
S^\om_{\mathcal{T}_i},\  \text{if}\ N\geq i,\\
*\ \text{if}\ N<i.
\end{cases}
\ee
One can think of $(\xi_n)_{n\geq 0}$ as the walk on the set $D$ visiting the trap locations in the same order as $\{S^\om_k :k \geq 0\}$.

\section{Main results} \label{sec:main_results}

As explained above, motivated by the dissipative sandpile model, our main aim is to introduce a trap landscape in which, as a function of a parameter, there is a transition between $\E(\tau)<\infty$
and $\E(\tau)=\infty$ (cf. \eqref{eq:survivaltime}).
It turns out that the appropriate growth of intervals between 
traps has to follow the recursive relation
\be
\label{eq:recursion}
    \forall j \geq 1: |I_{j+1}| = c|I_j|^2,
\ee
for some $c> |I_1|^{-1}$. The requirement $c> |I_1|^{-1}$ implies that $j \mapsto |I_j|$ is monotone increasing. The growth described by
\eqref{eq:recursion} is super-exponential, more precisely,
\be\label{eq:recusol}
|I_n|= c^{2^{n-1}-1} |I_1|^{2^{n-1}}
=\frac1c e^{2^{n-1}(\log (c)+ \log |I_1|)}.
\ee
We implicitly assume that $c$
is such that $|I_n|, n\in\N,$ is a sequence of integers. 

\br \label{rmk:possible_c}
 Using the expression in \eqref{eq:recusol}, we have the restriction
\be\label{restriction}
|I_{n+1}|= c^{2^{n}-1} |I_1|^{2^{n}} = (c|I_1|)^{2^n-1} |I_1| \in \N, \quad \text{for all } n \in \N. 
\ee
As we prove in the Appendix, this is only possible if
\be \label{eq:int_resriction}
c|I_1| \in \N, \quad \text{or equivalently,}\quad c = \frac{m}{|I_1|} \quad \text{for some } m \in \N. 
\ee
\er 

Our main result is then the following.
\bt \label{thm:main_result}
For $c>|I_1|^{-1}$ and traps distributed according to the recursion $|I_{j+1}| = c|I_j|^2$ we have
\begin{itemize}
    \item[1.] $\E(\tau) < \infty$ if     $c <1$. 
    \item[2.] $\E(\tau) = \infty$ if $c>1$.
\end{itemize}
\et

Theorem \ref{thm:main_result} is derived from two intermediate results. The first result, Theorem \ref{thm:criticality}, gives upper and lower bounds depending on the length of the first interval $|I_1|$. The second result, Theorem \ref{thm:finiteness_tail}, shows that in fact the finiteness of the expectation of the survival time only depends on the \emph{tail of the landscape $\om$}.  

\bt \label{thm:criticality}
Assume $|I_1|\geq 3$. For $c>|I_1|^{-1}$ and traps distributed according to the recursion $|I_{j+1}| = c|I_j|^2$ we have
\begin{itemize}
    \item[1.] $\E(\tau) < \infty$ if 
    \be\label{eq:c<}
    c<\frac{|I_1| + \sqrt{|I_1|^2-8}}{2 |I_1|}.
    \ee
    \item[2.] $\E(\tau) = \infty$ if
    \be\label{eq:c>}
    c > 1 + \frac{3}{|I_1|}.
    \ee
\end{itemize}
\et

Theorem \ref{thm:finiteness_tail} states that landscapes which are \emph{similar in the tail} have either both finite or both infinite expectation for the associated survival time. We say that two landscapes $\om^{(a)}$ and $\om^{(b)}$ (that don't necessarily satisfy the recursion in \eqref{eq:recursion}) are \textit{similar in the tail} if there exist $p_a,p_b \in \N$ such that

\be
\om^{(a)}(p_a + x) = \om^{(b)}(p_b + x)\qquad\hbox{for all }x \in \N.
\ee
To distinguish between the quantities corresponding to $\om^{(a)}$ and $\om^{(b)}$, we add superscripts $(a)$ and $(b)$, e.g. the survival time associated to $\om^{(a)}$ is denoted $\tau^{(a)}$
and the intervals between traps are denoted $I^{(a)}_i, i \geq 1$, whereas the survival time associated to $\om^{(b)}$ is denoted $\tau^{(b)}$ and the intervals between traps are denoted $I^{(b)}_i, i \geq 1$.

\bt \label{thm:finiteness_tail}
Let $\om^{(a)}$ and $\om^{(b)}$ be trap landscapes which are similar in the tail, then
\be
\E[\tau^{(a)}] < \infty \qquad \text{if and only if} \qquad \E[\tau^{(b)}] < \infty.
\ee
\et
Theorem \ref{thm:criticality} will be proved in Section \ref{sec:proofofmain} and Theorem \ref{thm:finiteness_tail} in Section \ref{subsec:proof_finiteness_tail}.\\

\subsubsection*{Proof of Theorem \ref{thm:main_result}}
We now argue that Theorem \ref{thm:main_result} follows from Theorem \ref{thm:criticality} and Theorem \ref{thm:finiteness_tail}.\\
Consider a landscape of traps $\om^{(a)}$ which satisfies the recursion in \eqref{eq:recursion} for some constant $c$. We now construct a new landscape $\om^{(b)}$.  To this end, we fix an arbitrary $k \in \N\setminus \{0\}$ and define 
\be\label{eq:new-interval}
|I_1^{(b)}| = |I_k^{(a)}| \qquad \text{and} \qquad |I_{i+1}^{(b)}| = c |I_i^{(b)}|^2 \text{ for all } i \in \N.
\ee
In other words, we construct $\om^{(b)}$ via the recursion in \eqref{eq:recursion} with the same constant $c$ as was used for the construction of $\om^{(a)}$. However, instead of starting the recursion with interval length $|I_1^{(a)}|$, we start from $|I_k^{(a)}|$. Notice that this construction is such that $\om^{(a)}$ and $\om^{(b)}$ are similar in the tail: for $p_a = x^{(a)}_{k-1}$ and $p_b = x^{(b)}_0 = 0$ we have
\[ 
\om^{(a)}( p_a + x) = \om^{(a)}( x^{(a)}_{k-1} + x) = \om^{(b)}(x) = \om^{(b)}( p_b + x).
\]
Recall that $k$ is an arbitrary natural number different from zero.  We can use Theorem \ref{thm:finiteness_tail} then Theorem \ref{thm:criticality} to conclude that for all $k \in \N\setminus\{0\}$ we must have
\begin{itemize}
    \item[1.] $\E(\tau^{(a)}) < \infty$ if 
    \[
    c<\frac{|I^{(a)}_k| + \sqrt{|I^{(a)}_k|^2-8}}{2 |I^{(a)}_k|}.
    \]\item[2.] $\E(\tau^{(a)}) = \infty$ if
    \[
    c > 1 + \frac{3}{|I^{(a)}_k|}.
    \]
\end{itemize} 
 This proves Theorem \ref{thm:main_result}, because $|I^{(a)}_k|\uparrow\infty$ as $k\to \infty$ (see \eqref{eq:recusol}) and
\[
\frac{y + \sqrt{y^2-8}}{2 y} \uparrow 1, \qquad  1 + \frac{3}{y} \downarrow 1 \qquad \text{as } y\rightarrow \infty.
\] 
\qed

Apart from Theorem \ref{thm:main_result}, this paper also contains an example for the \emph{non-monotonicity} of the expectation of the survival time as a function of the interval lengths. Consider the setting introduced in Section \ref{sec:Random_walk_in_a_landscape_of_soft_traps}. It is tempting to think that the expectation $\E(\tau)$ of the survival time
is monotonously increasing as a function of the interval lengths $|I_i|, i \geq 1$. That is, if the distances between the traps are all bigger in one landscape than in another, then the expectation of the associated survival time is bigger for the first landscape than for the second one. In Section \ref{subsec:counterexample} we show by an explicit counterexample that this intuition is in fact not correct. The tools we develop along the way will also be used to prove Theorem \ref{thm:finiteness_tail}. For this reason we included the proof of Theorem \ref{thm:finiteness_tail} in Section \ref{subsec:proof_finiteness_tail} at the end of Section \ref{sec:Monotonicity}.

\section{Proof of Theorem \ref{thm:criticality}}\label{sec:proofofmain}

Our strategy to prove Theorem \ref{thm:criticality} is the following. 
Recall that in \eqref{eq:N}  we defined 
$N = \sum_{j=0}^\infty \mathbb{I}(S^\omega_j \in D)$,
 the total number of visits of the 
set $D$ (see \eqref{eq:D}) before the random walk is trapped. 
Because at each visit of $D$ the random walk 
is trapped with probability $1/3$, $N$ is finite almost surely.
Recall that we defined  in \eqref{eq:Yi} the inter-arrival times as
$Y_i:= \mathcal{T}_i - \mathcal{T}_{i-1}$ for 
$ 1 \leq i \leq N$, and as $Y_i = 0$ for $i > N$. 
We have
\be\label{Etau-1}
\E(\tau) = \E\left( \sum_{i=1
}^N Y_i \right) = \E\left( \sum_{i=1
}^\infty  Y_i \right) = \sum_{i=1
}^\infty \E\left(  Y_i \right).
\ee
Our aim is then to derive estimates for $\E(\tau)$ by estimating the quantity 
$\E(Y_i)$  for a given $i\geq 1$.
To this end, we want to condition on the path of $(\xi_n)_{n \geq 0}$
up to time  $\mathcal{T}_{i-1}$. 
To facilitate the calculations we introduce notation 
for the events describing 
the increments of $(\xi_n)_{n \geq 0}$. 
The embedded random walk $\{\xi_l: l\in\N\}$ moves from a trap $x_i$ 
(for $i\geq 0$) towards the closest trap on its left ($x_{i-1}$) for $i>0$ 
or on its right ($x_{i+1}$) for $i\geq 0$, or remains at $x_i$.\\
This can be encoded by a sequence 
$\kappa\in \{-1,0,1\}^{i-1}$ for $i\geq 2$, where
$\kappa_l=1, -1$ means moving to the right, resp. left, whereas 
$\kappa_l=0$ means staying at the same trap location.
Since we are interested in estimating $\E(Y_i)$, we will be interested in the history of the embedded walk prior to time $i$, 
which explains why we chose the length $i-1$ for the sequence $\kappa$.\\
Fix $i \geq 2$ and let $\kappa \in \{-1,0,1\}^{i-1}$. Define the events 
\begin{eqnarray}\label{def:Akap}
    A_\kappa &:=& \{ \forall j \in\{1,...,i-1\}: \text{sgn}(\xi_{j} - \xi_{j-1}) = \kappa_j \},\\\label{def:Akaplus}
    A_\kappa^{+} &:=& A_\kappa
    \cap \{S_{\mathcal{T}_{i-1} + 1} = S_{\mathcal{T}_{i-1}} + 1 \}.
\end{eqnarray} 
Notice that $A_\kappa\subset (N\geq i)$, i.e., it is implicit in the definition of $A_\kappa$ that the embedded random walk hits (at least) $i$ traps before it reaches $*$.  One can interpret the event $A_\kappa$ as $(S_k)_{k\geq0}$ visiting the traps according to $\kappa$. We omit $i$ from our notation of $\kappa$ as well as of $A_\kappa$, for the sake of readability.

For example for $i=4$, $\kappa = (1,0,-1)$, dictates that 
$S_{\mathcal{T}_0} = x_0=0$, 
$S_{\mathcal{T}_1} = x_1$, $S_{\mathcal{T}_2} = x_1$ and 
$S_{\mathcal{T}_3} = x_0$, i.e.  the walker starts at $x_0$ then visits 
the trap at $x_1$ because $\kappa_1 = 1$, next it returns to $x_1$ 
before hitting any other trap because $\kappa_2 = 0$, and finally it 
moves back to $x_0$ because $\kappa_3 = -1$. 

In the subset  $A_\kappa^+$ of $A_\kappa$, the random walker jumps to the right 
after walking according to $\kappa$. Notice that not every 
$\kappa \in \{-1,0,1\}^{i-1}$ is compatible with the unilateral random walk, 
indeed, if $\sum_{j=1}^{l}\kappa_j<0$ for some 
$1\leq l\leq i-1$ then the corresponding walk would visit the left of 
the origin, which is not allowed (cf. \eqref{eq:transprob}). Therefore, 
we restrict $\kappa$ to the set
\be \label{eq:def_mathcalK}
\mathscr{K}:= \Big\{\kappa \in \{-1,0,1\}^{i-1}: \sum_{n=1}^{m} \kappa_n \geq 0 \text{ for all } m \in \{1,...,i-1\} \Big\}.
\ee
We also omit $i$ from our notation of $\mathscr{K}$.

For $i=1$ we define by convention $\mathcal{K} = \{()\}$ to be the set containing only the empty tuple, and correspondingly $A_\kappa=A_{()}$ and $A^+_\kappa=A_{()}^+$ are defined as the entire sample space, and therefore have probability one.

To prove Theorem \ref{thm:criticality}, the expectation $\E(\tau)$ in \eqref{Etau-1} is expanded in a sum over the contributions of all $\kappa \in \mathscr{K}$,
\be\label{Etau-2}
    \E(\tau) 
   = \sum_{i=1}^\infty \E\left(  Y_i \right)\\
    = \sum_{i=1}^\infty \sum_{\kappa \in \mathcal{K}}\Pr(A_\kappa) \E\left(   Y_i | A_\kappa  \right).
\ee

\subsection{Upper and lower bounds for inter-arrival times}\label{subsec:upper-lower}

The following lemma is essential in the proof of Theorem \ref{thm:criticality}. It provides an estimate for the contribution of each set $A_\kappa$ to $\E(\tau)$
(cf. \eqref{Etau-2}), comparing it to the contribution when $\kappa=\mathds{1}$,
where $\mathds{1}$ denotes the sequence where all $\kappa_l=1$ for $l=1, \ldots, i-1$.
Finally, it provides an explicit formula for the expected inter-arrival times for the case $\kappa=\mathds{1}$. The latter is also the main motivation why we chose the recursion $|I_{k+1}|= c|I_k|^2$.
To make the proof of this lemma more transparent and readable we provide an explicit Example \ref{ex:example_lemma}, in which some of the expressions below are given for a specific landscape of traps and a specific sequence $\kappa$. Example \ref{ex:example_lemma} is meant to be read alongside the proof of Lemma \ref{lemma:contributions}. 

\bl
\label{lemma:contributions}
Let $\kappa \in \mathscr{K}$. For traps distributed according to the recursion $|I_{j+1}| = c |I_j|^2$, $c \geq |I_1|^{-1}$ we have the following.
\ben
\item Upper bound for inter-arrival times:
\be
\label{eq:kappa_ubd}
\Pr ( A_{\kappa}) \E(Y_i|A_{\kappa}^+) \leq \Big( \frac{2}{c}\Big)^L \Big( \frac{2}{|I_1|^2c^2}\Big)^J \Pr ( A_{\mathds{1}}) \E(Y_i|A_\mathds{1}^+),
\ee
where $L = \sum_{n=1}^{i-1}\mathbb{I}(\kappa_n = 0)$ and  
$J = \sum_{n=1}^{i-1}\mathbb{I}(\kappa_n = -1)$.
\item Lower bound for inter-arrival times:
 If $\kappa \geq 0$, i.e. $ \kappa \in \{0,1\}^{i-1}$,
\be
\label{eq:kappa_lbd}
\Pr ( A_{\kappa}) \E(Y_i|A_{\kappa}^+) \geq \frac{1}{2} \Big( \frac{2}{c}\Big(1- \frac{1}{|I_1|}\Big)^2\Big)^L \Pr ( A_{\mathds{1}}) \E(Y_i|A_\mathds{1}^+).
\ee
\item Explicit formula for the case $\kappa=\mathds{1}$: 
\be\label{62}
\Pr(A_\mathds{1}) \E(Y_i|A_\mathds{1}^+)=  
\begin{cases}
\displaystyle{
 2|I_1|\left(\frac{c}{3}\right)^{i-1} \left(1-\frac{1}{|I_{i}|}\right)} \ 
\text{if}\ i\geq 2,
\\ \\
|I_1|-1 \ \text{if}\ i=1
\end{cases}.
\ee
As a consequence
\be\label{62-again}
\Pr(A_\mathds{1}) \E(Y_i|A_\mathds{1}^+)  \leq 2\left(\frac{c}{3}\right)^{i-1} |I_1|.
\ee
\een
\el

\bpr

\noindent
\textbf{Step 1.}
We first provide exact expressions for the transition probabilities of $\{\xi_l: l\in \N \}$, which we will use extensively. For two traps $x_i, x_j \in D$:

\be \label{eq:transition_emb}
\Pr(\xi_{l+1} = x_j| \xi_{l} = x_i) = 
\begin{cases}
    \displaystyle{\frac{2}{3}\left(\frac{1}{2|I_i|} \right)}\quad &\text{if } i \geq 1 \text{ and } j = i-1,\\  \\
    \displaystyle{\frac{2}{3} \left( \frac{2^{\mathbb{I}(i=0)}}{2|I_{i+1}|}\right)} \quad &\text{if } j = i+1,\\ \\
    \displaystyle{\frac{2}{3}\left(1 - \frac{1}{2|I_{i+\mathbb{I}(i=0)}|} - \frac{1}{2|I_{i+1}|}\right) }\quad &\text{if } j=i,\\  \\
    0 \quad &\text{otherwise.}
\end{cases}.
\ee
Notice that 
$\sum_{y\in D}\Pr(\xi_{l+1} = y| \xi_{l} = x_i)=2/3$, which corresponds to the fact that upon hitting the trap $x_i$, the walker survives with probability $2/3$ and then moves to the next trap.

The transition probabilities in \eqref{eq:transition_emb} derive from the fact that for $\alpha < \beta < \gamma \in \N$, the probability the (not trapped) random walk  $\{S_n: n \in \N \}$ reaches $\gamma$ before it reaches $\alpha$ starting from $\beta$ is equal to 

\be \label{eq:cross_prob_interval}
\frac{\gamma-\beta}{\gamma-\alpha}.
\ee
Indeed,
\begin{itemize}
    \item[(i)] The probability that the embedded random walk makes a transition from $x_i$ to the first trap on its left, $x_{i-1}$, is equal to the probability that the walk  does not get trapped at $x_i$, next, it moves from $x_i$ one step to the left (i.e., to $x_{i}-1$) and then it reaches $x_{i-1}$ without returning to $x_i$ first. Not getting trapped on $x_i$ happens with probability $2/3$. The step to the left occurs with probability $1/2$. Finally by \eqref{eq:cross_prob_interval},
    reaching $x_{i-1}$ without returning to $x_i$ first happens with probability $1/ |I_i|$ (recall from \eqref{eq:Ii} that $I_i= [x_{i-1}, x_i)$, so $|I_i| = x_i-x_{i-1}$).

    \item[(ii)] The transition probability from $x_i$ to the first trap on the right, $x_{i+1}$, can be found in the same way: If $i \neq0$ the random walker survives with probability $2/3$ and moves one step to the right with probability $1/2$ and then reaches $x_{i+1}$ before it returns to $x_i$ with probability $1/|I_{i+1}|$. In the case where $i = 0$, the  probability of jumping right is one instead of $1/2$, hence there is a correction factor $2^{\mathbb{I}(i=0)}$.

    \item[(iii)] The transition probability from $x_i$ back to $x_i$ is simply the probability of survival multiplied by one minus the probability that the walker jumps either to the left or to the right trap before returning to $x_i$. In the case where $i=0$, the walker can only jump to the right from $x_i$. Hence the transition probability is $(2/3)(1-|I_1|^{-1})$.
\end{itemize}

\noindent
\textbf{Step 2.}
As already explained, the idea of the proof is the following. For a given arbitrary $\kappa \in \mathcal{K}$, we want to find bounds 
on $\Pr ( A_{\kappa}) \E(Y_i|A_{\kappa}^+)$ in terms of the quantity
$ \Pr ( A_{\mathds{1}}) \E(Y_i|A_\mathds{1}^+).$
Our strategy is therefore to estimate how this quantity changes when we modify the entries of $\mathds{1} = (1,...,1) \in \N^{i-1}$ one by one from left to right. In this way, starting from the sequence $\mathds{1}=(1,...,1)$ one can construct each sequence $\kappa \in \mathcal{K}$. However, each time an entry with value 1 is replaced by -1 or 0, a correction factor has to be included in the estimates. This explains the appearance of the powers $J$ and $L$, the number of minus ones and zeros respectively, in the estimates  \eqref{eq:kappa_ubd} and \eqref{eq:kappa_lbd}.

Suppose that in the process of changing from the sequence
$\mathds{1}=(1,...,1)$ to the sequence $\kappa$ we have arrived at entry $s-1$ and we want to alter entry $s$ in the next step. That is, we consider a sequence $\kappa$ for which there exists an $ s \in \{1,...,i-1\}$ such that for all $j \geq s: \kappa_j = 1$. For now assume that $s\neq1$. In other words, we are not changing the first entry of $\kappa$. The case where $s = 1$, where we do change the first entry, will be treated soon after. Let $[\kappa]_{1}^{s-1} := (\kappa_1, \dots,\kappa_{s-1})$ denote the sequence $\kappa$ restricted to entries between $1$ and $s-1$ and let 
\be\label{eq:Theta}\Theta = \Theta(\kappa,s) = \sum_{j = 1}^{s-1} \kappa_j\ee
 denote the trap number of the position of the embedded random walk after following the steps prescribed by $[\kappa]_1^{s-1}$. Using the transition probabilities in \eqref{eq:transition_emb} we find

\be \label{eq:PA_kappa}
    \Pr (A_\kappa) = \Pr(A_{[\kappa]_{1}^{s-1}}) \cdot 2^{\mathbb{I}(\Theta = 0)} \cdot \prod_{j = \Theta + 1}^{\Theta + i-s} \frac{2}{3} \left(\frac{1}{2|I_j|}\right).
\ee
The factor $\Pr(A_{[\kappa]_{1}^{s-1}})$ comes from the embedded walk following $\kappa$ up to entry $s-1$. After doing so, the walker is at the trap with number $\Theta$. The remaining entries of $\kappa$ are all one and there are $(i-1)-(s-1) = i-s$ of them. This explains the product over $j$ in \eqref{eq:PA_kappa}. In the case where $\Theta = 0$ a factor two must be added to account for the fact that the first step of the walker leaving $x_0$ is always to the right (cf. the second line in \eqref{eq:transition_emb}). In Example \ref{ex:example_lemma} equation \eqref{eq:PA_kappa} corresponds to equation \eqref{eq:PA_kappa_example}.\\
We now come back to the case where $s = 1$. In this scenario the sequence $[\kappa]_{1}^{s-1}$ will be defined as the empty tuple: $[\kappa]_{1}^{0} := ()$. According to our earlier conventions for $A_{()}$ and $A_{()}^+$ we obtain

\be \label{eq:def_s=1}
 \Pr(A_{[\kappa]_{1}^0}) = \Pr(A_{()}) = 1.
\ee

Recall that $s$ is such that for all $j \geq s: \kappa_j = 1$, so for $s=1$, $\kappa$ must be the all one vector $\mathds{1} = (1,1,\dots,1) \in \{-1,0,1\}^{i-1}$. Also notice that $\Theta = \sum_{j=1}^{s-1} \kappa_j = 0$ since the sum is empty. We observe that equation \eqref{eq:def_s=1} implies that the right hand side of Equation \eqref{eq:PA_kappa}  becomes the probability associated to $A_\mathds{1}$:

\be\label{eq:PA1}
    \Pr (A_\kappa) =  2 \prod_{j =  1}^{i-1} \frac{2}{3} \left(\frac{1}{2|I_j|}\right) = \Pr(A_{\mathds{1}}).
\ee
Let 
\be\label{eq:fthetais}
f = f(\Theta,i,s) = \Theta + i - s + 1.
\ee
Then $x_{f-1}$ is the position where the random walk ends up after following $\kappa$, and $I_f= [x_{f-1}, x_f)$ is the corresponding interval to the right of $x_{f-1}$. Notice that the last interval crossed by the random walk following $\kappa$ is $I_{f-1}$. For the expectation $\E(Y_i|A_{\kappa}^+)$ we claim that

\be \label{eq:EY_kappa}
\E(Y_i|A_{\kappa}^+) =  |I_f|-1.
\ee
To see this we use the following well known fact of simple random walk. For $0\leq \alpha < \beta < \gamma$, the expectation of the time before hitting either $\alpha$ or $\gamma$, starting from $\beta$, is given by

\be \label{eq:time_in_interval}
(\beta-\alpha) \cdot (\gamma-\beta).
\ee
The event $A_{\kappa}^+$ corresponds to the walker walking a trajectory consistent with $\kappa$ after which it survives the trap $x_{f-1}$ and jumps one step to the right, i.e., to $x_{f-1}+1\in I_{f}$. At this point the walker must have survived precisely $i-1$ traps. Recall (cf. \eqref{eq:Yi}) that $Y_i$ is the time interval between hitting a trap for the $(i-1)$-th and the $i$-th time. Therefore, $Y_i$ is precisely the time it takes to hit either $x_{f-1}$ or $x_f$ starting from position $x_{f-1}+1$. Due to \eqref{eq:time_in_interval}, we indeed have $\E(Y_i|A_{\kappa}^+) = x_f - x_{f-1} -1=  |I_f|-1$.

We now see how to alter entry $s$.
Let $\sigma \in \{-1,0\}$. We define $\kappa^{s,\sigma}$ as the sequence $\kappa$ where the entry number $s \in \{ 1, \dots, i-1\}$ is replaced by $\sigma$:

\be\label{eq:kappassigma}
(\kappa^{s,\sigma})_j = 
    \begin{cases}
    \sigma \quad \text{if } j = s\\
    \kappa_j \quad \text{if }j \neq s
    \end{cases}.
\ee
We then analyse the differences between the quantities $\Pr ( A_{\kappa}) \E(Y_i|A_\kappa^+)$
and $\Pr ( A_{\kappa^{s,\sigma}}) \E(Y_i|A_{\kappa^{s,\sigma}}^+)$ according to the value of $\sigma$.

\subsubsection*{Replacing entry number $s$ by $\mathbf{\sigma = 0}$}

Recall formula \eqref{eq:PA_kappa}, and the comments about this formula. We now write it in a way suitable for what follows, that is, by looking separately at the two possibilities for $s$:

\be \label{eq:PA_kappa-2}
    \Pr (A_\kappa) = \Pr(A_{[\kappa]_{1}^{s-1}}) \cdot\left[ 2^{\mathbb{I}(\Theta = 0)}\cdot\frac{2}{3} \left(\frac{1}{2|I_{\Theta + 1}|}\right) \right]\cdot
    \left\{\mathbb{I}(s=i-1) + \mathbb{I}(s<i-1) \prod_{j = \Theta + 2}^{\Theta + i-s} \frac{2}{3} \left(\frac{1}{2|I_j|}\right)\right\}.
\ee
The last terms come from the fact that either $s=i-1$ and there is no other term to add after the previous ones, or $s<i-1$ and there is a product of terms to add  to go from entry $s$ to entry $i-1$ in $\kappa$; these terms are similar to the second term (apart from $2^{\mathbb{I}(\Theta = 0)}$).\\

Then 
$\Pr(A_{\kappa^{s,0}})$ can be expressed as
follows: 
\beq\label{eq:PA_sig=0}
    \Pr (A_{\kappa^{s,0}}) 
 &=& \Pr(A_{[\kappa]_{1}^{s-1}})\cdot \frac{2}{3}\Big(1-\frac{1}{2|I_{\Theta+1}|}-\frac{1}{2|I_{\Theta + \mathbb{I}(\Theta = 0)}|}\Big) 
 \times\\ \nonumber 
 &&\qquad\left\{\mathbb{I}(s=i-1) + \mathbb{I}(s<i-1)
 2^{\mathbb{I}(\Theta = 0)} \cdot \prod_{j = \Theta + 1}^{\Theta + i-s-1} \frac{2}{3}\left(\frac{1}{2|I_j|}\right)\right\}.\eeq
The first factor $\Pr(A_{[\kappa]_{1}^{s-1}})$ is the same as in \eqref{eq:PA_kappa-2} because the  entries $1$ to $s-1$ are identical for $\kappa$ and $\kappa^{s,0}$. After following $\kappa^{s,0}$ up to entry $s-1$, the embedded walk will be at trap $x_\Theta$, to which it returns in the next step since $(\kappa^{s,0})_s = 0$. This gives the second factor (cf. the third line in \eqref{eq:transition_emb}). The third factor corresponds to the steps left  to go from $s$ to $i-1$. So, as in \eqref{eq:PA_kappa-2}, either $s=i-1$ and we are done, or $s<i-1$ and there still is a tail of ones in $\kappa^{s,0}$; this tail of ones  is precisely one shorter than the tail of ones in $\kappa$. Hence the product goes only up to $\Theta + i-s-1$ instead of up to $\Theta + i-s$, as it does in \eqref{eq:PA_kappa-2}. Note also that the product begins at $\Theta+1$ instead of $\Theta+2$, since  at step $s$ we are now on $x_\Theta$; we therefore have to add the factor $2^{\mathbb{I}(\Theta = 0)}$ for this first term of the product.\\

Equation \eqref{eq:PA_sig=0} corresponds to \eqref{eq:PA_kappa_0_ex} in Example \ref{ex:example_lemma}. Note that in this example we have $s<i-2$. \\

We combine  \eqref{eq:PA_kappa-2} and \eqref{eq:PA_sig=0} to obtain, recall that $f - 1 = \Theta +i - s $ (cf. \eqref{eq:fthetais}),

 \beq\label{eq:PA_sig=0_relation}
 \Pr (A_{\kappa^{s,0}}) 
 &=& \mathbb{I}(s=i-1) \frac{2}{3}\Big(1-\frac{1}{2|I_{\Theta+1}|}-\frac{1}{2|I_{\Theta + \mathbb{I}(\Theta = 0)}|}\Big) 
 \Pr(A_{\kappa})\cdot\frac{3}{2}\cdot\frac{2|I_{\Theta + 1}|}{2^{\mathbb{I}(\Theta = 0)}}\\  \nonumber
 &&\, + \mathbb{I}(s<i-1)
 \frac{2}{3}\Big(1-\frac{1}{2|I_{\Theta+1}|}-\frac{1}{2|I_{\Theta + \mathbb{I}(\Theta = 0)}|}\Big) 
 \Pr(A_{\kappa})\cdot\frac{3}{2}\cdot 2|I_{\Theta+i-s}|
 \\ \nonumber
 &=& \Pr(A_{\kappa})\Big(1-\frac{1}{2|I_{\Theta+1}|}-\frac{1}{2|I_{\Theta + \mathbb{I}(\Theta = 0)}|}\Big)
 \cdot 2|I_{f-1}|\left\{\frac{\mathbb{I}(s=i-1)}{2^{\mathbb{I}(\Theta = 0)}}
  + \mathbb{I}(s<i-1)\right\}
\eeq

Notice that 
\be\label{simpler}
2\left\{\frac{\mathbb{I}(s=i-1)}{2}
+ \mathbb{I}(s < i-1)\right\}
= 2^{\mathbb{I}(s < i-1)}.
\ee

 Thus we have, combining \eqref{eq:PA_sig=0_relation}  and \eqref{simpler},
 upper and lower bounds for the probability $\Pr (A_{\kappa^{s,0}})$ in terms of $\Pr(A_{\kappa})$, 

\be \label{eq:PA_final_ub}
     2^{\mathbb{I}(s < i-1)}|I_{f-1}|\Big(1-\frac{1}{|I_{1}|}\Big) \cdot \Pr (A_{\kappa}) \leq \Pr (A_{\kappa^{s,0}}) \leq 2 |I_{f-1}| \cdot \Pr (A_{\kappa}).
\ee
Here, for the lower bound, we used the lengths of the intervals increase in their index. Next we compare $\E(Y_i|A_{\kappa^{s,0}}^+)$ and $\E(Y_i|A_{\kappa}^+)$. We can use an argument similar to the one we used for \eqref{eq:EY_kappa} to see that $\E(Y_i|A_{\kappa^{s,0}}^+)$ is equal to the length of the interval to the right of the trap where the walker ends up after following $\kappa^{s,0}$. This interval is $I_{f-1}$, since the walker ends up at $x_{f-2}$, the trap directly to the left of $x_{f-1}$ (recall that $x_{f-1}$ is the trap where the walker ends up after walking according to $\kappa$). Using \eqref{eq:EY_kappa} we find that 

\be \label{eq:EY_sig=0}
\E(Y_i|A_{\kappa^{s,0}}^+) = |I_{f-1}| -1 = \frac{|I_{f-1}|-1}{|I_f|-1} \cdot (|I_f|-1) = \frac{|I_{f-1}|-1}{|I_f|-1} \cdot \E(Y_i|A_{\kappa}^+).
\ee
This equation corresponds to \eqref{eq:EY_sig=0_ex} in Example \ref{ex:example_lemma}. 

The relation in \eqref{eq:PA_sig=0_relation} gives us upper and lower bounds for the probability $\Pr (A_{\kappa^{s,0}})$ in terms of $\Pr(A_{\kappa})$,

\be \label{eq:PA_final_ub}
     2^{\mathbb{I}(s < i-1)}|I_{f-1}|\Big(1-\frac{1}{|I_{1}|}\Big) \cdot \Pr (A_{\kappa}) \leq \Pr (A_{\kappa^{s,0}}) \leq 2 |I_{f-1}| \cdot \Pr (A_{\kappa}).
\ee
Here, for the lower bound, we used the lengths of the intervals increase in their index. Next we compare $\E(Y_i|A_{\kappa^{s,0}}^+)$ and $\E(Y_i|A_{\kappa}^+)$. We can use an argument similar to the one we used for \eqref{eq:EY_kappa} to see that $\E(Y_i|A_{\kappa^{s,0}}^+)$ is equal to the length of the interval to the right of the trap where the walker ends up after following $\kappa^{s,0}$. This interval is $I_{f-1}$, since the walker ends up at $x_{f-2}$, the trap directly to the left of $x_{f-1}$ (recall that $x_{f-1}$ is the trap where the walker ends up after walking according to $\kappa$). Using \eqref{eq:EY_kappa} we find that 

\be \label{eq:EY_sig=0}
\E(Y_i|A_{\kappa^{s,0}}^+) = |I_{f-1}| -1 = \frac{|I_{f-1}|-1}{|I_f|-1} \cdot (|I_f|-1) = \frac{|I_{f-1}|-1}{|I_f|-1} \cdot \E(Y_i|A_{\kappa}^+).
\ee
This equation corresponds to \eqref{eq:EY_sig=0_ex} in Example \ref{ex:example_lemma}. Notice that, after \eqref{eq:EY_sig=0},

\begin{align} 
\E(Y_i|A_{\kappa^{s,0}}^+) 
&=  \left(\frac{(|I_{f-1}|-1)|I_f|}{(|I_f|-1)|I_{f-1}|}\right) \cdot \frac{|I_{f-1}|}{|I_f|} \cdot \E(Y_i|A_{\kappa}^+) \label{eq:EY_sig=0_{ub}}\\
&= \left( \frac{|I_{f-1}||I_f| -|I_f|}{|I_{f-1}||I_f| -|I_{f-1}|}\right) \cdot \frac{|I_{f-1}|}{|I_f|} \cdot \E(Y_i|A_{\kappa}^+) \leq   \frac{|I_{f-1}|}{|I_f|}\cdot \E(Y_i|A_{\kappa}^+). \nonumber
\end{align}
Finally we combine \eqref{eq:PA_final_ub} and \eqref{eq:EY_sig=0_{ub}} together with the recursion $|I_{j+1}| = c |I_j|^2$ to obtain the following bounds

\begin{align} \label{eq:final_ub_sig=0}
    \Pr(A_{\kappa^{s,0}}) \E(Y_i| A_{\kappa^{s,0}}^+) &\leq 2|I_{f-1}| \cdot \Pr(A_{\kappa}) \cdot  \frac{|I_{f-1}|}{|I_f|} \cdot \E(Y_i|A_{\kappa}^+)\\
    &= \frac{2}{c} \cdot \Pr(A_{\kappa}) \E(Y_i|A_{\kappa}^+)\nonumber
\end{align}
and
\begin{align}\label{eq:final_lb_sig=0}
    \Pr(A_{\kappa^{s,0}}) \E(Y_i| A_{\kappa^{s,0}}^+) &\geq 2^{\mathbb{I}( s < i-1) }|I_{f-1}|\Big(1-\frac{1}{|I_1|}\Big) \cdot \Pr(A_{\kappa}) \cdot \frac{|I_{f-1}|-1}{|I_f|-1} \cdot \E(Y_i|A_{\kappa}^+) \\
    &\geq 2^{\mathbb{I}( s < i-1) }\Big(1-\frac{1}{|I_1|}\Big) \cdot \Pr(A_{\kappa}) \cdot \frac{|I_{f-1}|(|I_{f-1}|-1)}{|I_f|} \cdot \E(Y_i|A_{\kappa}^+) \nonumber\\
    &= 2^{\mathbb{I}( s < i-1) }\Big(1-\frac{1}{|I_1|}\Big) \cdot \Pr(A_{\kappa}) \cdot \frac{1}{c} \left(1-\frac{1}{|I_{f-1}|}\right) \cdot \E(Y_i|A_{\kappa}^+) \nonumber\\
    &\geq \frac{2^{\mathbb{I}( s < i-1) }}{c} \Big(1-\frac{1}{|I_1|}\Big)^2 \cdot \Pr(A_{\kappa}) \cdot \E(Y_i|A_{\kappa}^+). \nonumber
\end{align}

\subsubsection*{Replacing entry number $s$ by $\mathbf{\sigma = -1}$}

In this part of the proof we assume that $\kappa$ is such that $\kappa^{s,-1}$ is in $\mathcal{K}$. This assumption implies that $s\geq 2$. Since $s\leq i-1$,
we also have $i\geq 3$. 

As previously, recall formula \eqref{eq:PA_kappa}, and the comments about this formula. We again write it in a way suitable for what follows, that is, by looking separately at the three possibilities for $s$:

\begin{align} \label{eq:PA_kappa-3}
    \Pr (A_\kappa) = &\Pr(A_{[\kappa]_{1}^{s-1}}) \cdot\left[ 2^{\mathbb{I}(\Theta = 0)}\cdot\frac{2}{3} \left(\frac{1}{2|I_{\Theta + 1}|}\right)\right]\\
    &\cdot \left\{\mathbb{I}(s=i-1) + \mathbb{I}(s=i-2)\frac{2}{3} \left(\frac{1}{2|I_{\Theta + 2}|}\right) + \mathbb{I}(s<i-2) \prod_{j = \Theta + 2}^{\Theta + i-s} \frac{2}{3} \left(\frac{1}{2|I_j|}\right)\right\}.\nonumber
\end{align}

The second term above corresponds to the entry $s$ in $\kappa$, according to the second line in formula \eqref{eq:transition_emb}. Note that in the case where $\Theta = 0$ a factor two must be added to account for the fact that the first step of the walker leaving $x_0$ is always to the right. Then either $s=i-1$ and there is no other term, or $s=i-2$ and there is one term for the step from $i-2$ to $i-1$ similar to the second term (apart from $2^{\mathbb{I}(\Theta = 0)}$), or $s<i-2$ and there is a product of terms similar to the second term (apart from $2^{\mathbb{I}(\Theta = 0)}$) to go from entry $s$ to entry $i-1$ in $\kappa$.\\

Then,  
$\Pr(A_{\kappa^{s,-1}})$ can be expressed as
follows. 

\beq \label{eq:PA_sig=-1}
    \Pr (A_{\kappa^{s,-1}}) 
 &=&  \Pr(A_{[\kappa]_{1}^{s-1}}) \left[\frac{2}{3}\cdot\frac{1}{2|I_{\Theta}|}\right]\cdot 
\left\{\mathbb{I}(s=i-1)+ \mathbb{I}(s=i-2)\cdot 2^{\mathbb{I}(\Theta = 1)}\frac{2}{3} \left(\frac{1}{2|I_\Theta|}\right)\right.\\ \nonumber &&\qquad+ \left.\mathbb{I}(s<i-2)\cdot
 2^{\mathbb{I}(\Theta = 1)} \cdot \prod_{j = \Theta}^{\Theta + i-s-2} \frac{2}{3} \left(\frac{1}{2|I_j|}\right)\right\}.
 \eeq

 The first factor is the same as in \eqref{eq:PA_kappa-3} because  the entries $1$ to $s-1$ are identical for $\kappa$ and $\kappa^{s,-1}$. After following $\kappa^{s,-1}$ up to entry $s-1$, the embedded walk is at trap $x_\Theta$, from which it will move to $x_{\Theta-1}$ since $(\kappa^{s,-1})_s = -1$. This gives the second factor (cf. the first line in formula (28)).  The third factor corresponds to the steps left  to go from $s$ to $i-1$. So, as in \eqref{eq:PA_kappa-3}, 
 there are 3 cases: either $s=i-1$ and we are done, or $s=i-2$  and there is one term for the step from $i-2$ to $i-1$, this step is to the right, according to the second line in formula \eqref{eq:transition_emb}, so that it requires the factor $2^{\mathbb{I}(\Theta -1= 0)}$; or $s<i-2$ and there still is a tail of ones in $\kappa^{s,-1}$, that  is precisely one shorter than the tail of ones in $\kappa$.
 Because at step $s$, the walker is at $x_{\Theta - 1}$, comparing to \eqref{eq:PA_kappa-3}, the product in \eqref{eq:PA_sig=-1} starts from $\Theta$ and goes up to $\Theta +i-s-2$ instead of starting at $\Theta+1$ and going up to $\Theta+i-s$, as it does in \eqref{eq:PA_kappa-3}. Equation \eqref{eq:PA_sig=-1} corresponds to \eqref{eq:PA_kappa_-1_ex} in Example \ref{ex:example_lemma}.
  Note that in this example we have $s<i-2$. \\

We combine \eqref{eq:PA_kappa-3} and \eqref{eq:PA_sig=-1} to obtain
 
 \begin{align}\label{eq:PA_sig=-1_relation}
 &\Pr (A_{\kappa^{s,-1}}) \\
 &= \mathbb{I}(s=i-1)\cdot \left(\frac{2}{3}\cdot\frac{1}{2|I_{\Theta}|} \right)^2\cdot \Pr(A_{\kappa})\cdot\left(\frac{3}{2}\cdot 2|I_{\Theta}| \right)\cdot\frac{3}{2}\cdot\frac{2|I_{\Theta + 1}|}{2^{\mathbb{I}(\Theta = 0)}}\nonumber \\  
 &\quad+ \mathbb{I}(s=i-2)\cdot 2^{\mathbb{I}(\Theta = 1)}\left(\frac{2}{3} \cdot\frac{1}{2|I_\Theta|}\right)^2\cdot \Pr(A_{\kappa})\cdot\frac{3}{2}\cdot\frac{2|I_{\Theta + 1}|}{2^{\mathbb{I}(\Theta = 0)}}\cdot\frac{3}{2}\cdot 2|I_{\Theta + 2}| \nonumber\\ 
 &\quad +\mathbb{I}(s<i-2) \cdot \frac{2}{3}\cdot\frac{1}{2|I_{\Theta}|} \cdot \Pr(A_{\kappa})\cdot\frac{3}{2}\cdot 2|I_{\Theta+i-s-1}|\cdot\frac{3}{2}\cdot 2|I_{\Theta+i-s}|\frac{2}{3} \left(\frac{2^{\mathbb{I}(\Theta = 1)}}{2|I_{\Theta}|}\right)\frac{2}{3} \left(\frac{1}{2|I_{\Theta+1}|}\right) \nonumber\\ 
 &= \mathbb{I}(s=i-1)\cdot \left(\frac{1}{|I_{\Theta}|} \right)^2\cdot \Pr(A_{\kappa})\cdot |I_{f-2}|\cdot\frac{|I_{f-1}|}{2^{\mathbb{I}(\Theta = 0)}}\nonumber\\  
 &\quad + \mathbb{I}(s=i-2)\cdot 2^{\mathbb{I}(\Theta = 1)}\left(\frac{1}{|I_\Theta|}\right)^2\cdot \Pr(A_{\kappa})\cdot\frac{|I_{f-2}|}{2^{\mathbb{I}(\Theta = 0)}}\cdot |I_{f-1}|\nonumber\\  
 &\quad +\mathbb{I}(s<i-2) \cdot 2^{\mathbb{I}(\Theta = 1)}\left(\frac{1}{|I_\Theta|}\right)^2\cdot \Pr(A_{\kappa})\cdot |I_{f-2}|\cdot |I_{f-1}| \frac{2}{3} \left(\frac{1}{2|I_{\Theta+1}|}\right)\nonumber\\ 
 &= \Pr(A_{\kappa})\cdot\left(\frac{1}{|I_\Theta|}\right)^2\cdot |I_{f-2}|\cdot |I_{f-1}|\nonumber\\ 
 &\quad\left\{\frac{\mathbb{I}(s=i-1)}{2^{\mathbb{I}(\Theta = 0)}} + \mathbb{I}(s=i-2)\cdot\frac{2^{\mathbb{I}(\Theta = 1)}}{2^{\mathbb{I}(\Theta = 0)}} + \mathbb{I}(s<i-2)\cdot 2^{\mathbb{I}(\Theta = 1)}\cdot  \frac{2}{3} \left(\frac{1}{2|I_{\Theta+1}|}\right)\right\}. \nonumber
\end{align}

In this computation,
because $f-1=\Theta+i-s$, for $s=i-1$ we have $f-1=\Theta+1$, while for $s=i-2$ we have $f-1=\Theta+2$. Thus we have

\begin{align}\label{other50}
\Pr (A_{\kappa^{s,-1}}) \leq&
\Pr(A_{\kappa}) \Big(\frac{1}{|I_{1}|}\Big)^2
|I_{f-2}|\cdot |I_{f-1}|\\
&\cdot\left\{\frac{\mathbb{I}(s=i-1)}{2} + 2\cdot\mathbb{I}(s=i-2) + \frac{2}{3}\cdot \mathbb{I}(s<i-2) \right\} \nonumber\\
\leq& \Pr(A_{\kappa}) \Big(\frac{1}{|I_{1}|}\Big)^2
2|I_{f-2}|\cdot |I_{f-1}|.\nonumber 
\end{align} 
Moreover, after the walker follows $\kappa^{s,-1}$, it ends up at $x_{f-3}$. Hence the interval to the right is $I_{f-2}$. This yields (recall \eqref{eq:EY_kappa})

\be \label{eq:EY_sig=-1}
\E(Y_i|A_{\kappa^{s,-1}}^+) = |I_{f-2}| -1 = \frac{|I_{f-2}|-1}{|I_f|-1} \cdot (|I_f|-1) = \frac{|I_{f-2}|-1}{|I_f|-1} \cdot \E(Y_i|A_{\kappa}^+),
\ee
which can be bounded from above by (cf. \eqref{eq:EY_sig=0_{ub}})
\begin{align} \label{eq:EY_sig=-1_{ub}}
\E(Y_i|A_{\kappa^{s,-1}}^+) 
\leq   \frac{|I_{f-2}|}{|I_f|}\cdot \E(Y_i|A_{\kappa}^+).
\end{align}

This equation corresponds to \eqref{eq:EY_sig=-1_ex} in Example \ref{ex:example_lemma}. We can use the recursion $|I_{j+1}| = c|I_j|^2$ to find the following upper bound, 

\begin{align} \label{eq:final_ub_sig=-1}
    \Pr(A_{\kappa^{s,-1}}) \E(Y_i| A_{\kappa^{s,-1}}^+) &\leq \frac{2}{|I_1|^2} \cdot |I_{f-1}||I_{f-2}|\cdot \Pr(A_{\kappa}) \cdot  \frac{|I_{f-2}|}{|I_f|} \cdot \E(Y_i|A_{\kappa}^+)\\
    &= \frac{2}{(|I_1|c)^2} \cdot \Pr(A_{\kappa}) \E(Y_i|A_{\kappa}^+). \nonumber
\end{align}

\subsubsection*{Conclusion of the proof of \eqref{eq:kappa_ubd} and \eqref{eq:kappa_lbd}}
By replacing the entries in $\mathds{1} = (1,...,1) \in \N^{i-1}$ from left to right we can construct any vector $\kappa \in \{-1,0,1\}^{i-1}$ which corresponds to a valid path of the embedded random walk. To prove the upper bound \eqref{eq:kappa_ubd} we use equations \eqref{eq:final_ub_sig=0} and \eqref{eq:final_ub_sig=-1} repeatedly. The upper bound then gains a factor $2c^{-1}$ for each time a zero is added. Likewise it gains a factor $2(c|I_1|)^{-2}$ for every $-1$. Equation \eqref{eq:kappa_ubd} follows. One can understand equation \eqref{eq:kappa_lbd} in the same way. Each time a one is replaced by a zero, the factor $2c^{-1}(1-|I_1|^{-1})$ is added (see \eqref{eq:final_lb_sig=0}). Moreover in case also the last entry of $\kappa$ is zero, a correction of a factor $1/2$ is needed, because of the term $2^{\mathbb{I}( s < i-1) }$ in \eqref{eq:final_lb_sig=0}.

\subsubsection*{Proof of \eqref{62}}

\noindent
Here we calculate the factor $\Pr(A_\mathds{1}) \E(Y_i|A_\mathds{1}^+)$. 

For $i=1$, the sequence $\kappa$ is empty and hence

\be \label{eq:case_i=1}
\Pr(A_{\kappa}) = 1, \qquad \E(Y_1|A_\kappa^+) = \E(Y_1) = |I_1|-1, \quad \Pr(A_{\kappa}) \E(Y_1|A_\kappa^+) = |I_1|-1. 
\ee
For $i\geq 2$ we  see that,  using \eqref{eq:PA1} and \eqref{eq:EY_kappa},

\be \label{eq:formula_P1E1}
 \Pr(A_\mathds{1}) \E(Y_i|A_\mathds{1}^+) =2 \prod_{j=1}^{i-1} \left( \frac{2}{3} \cdot \frac{1}{ 2 |I_j|} \right) \cdot \left(|I_{i}|-1\right).
\ee 
By induction we show that 

\be \label{eq:induction_claim}
    2 \prod_{j=1}^{i-1} \left( \frac{2}{3} \cdot \frac{1}{ 2 |I_j|} \right) \cdot |I_{i}| = 2 \left(\frac{c}{3}\right)^{i-1} |I_1|.
\ee
For $i = 2$ we indeed have

\[
2 \prod_{j=1}^{i-1} \left( \frac{2}{3} \cdot \frac{1}{ 2 |I_j|} \right) \cdot |I_{i}| = \frac{2}{3} \frac{|I_2|}{|I_1|} = \frac{2}{3} \frac{c |I_1|^2}{|I_1|} = 2\left(\frac{c}{3}\right)^1 |I_1|.
\]
Now assume that \eqref{eq:induction_claim} holds for some $i = k \geq 2$, then

\begin{align}
\left( 2 \prod_{j=1}^{k} \frac{1}{ 3 |I_j|} \right)  |I_{k+1}| &= \left( 2 \prod_{j=1}^{k-1} \frac{1}{ 3 |I_j|} \right) \left(\frac{1}{3|I_k|}\right)  c|I_k|^2\\
&=\left(\frac{c}{3}\right) \left( 2 \prod_{j=1}^{k-1} \frac{1}{3 |I_j|} \right)  |I_{k}| = 2\left(\frac{c}{3}\right)^k |I_1|. \nonumber
\end{align}
This proves \eqref{eq:induction_claim}. 

 Putting together \eqref{eq:formula_P1E1} and \eqref{eq:induction_claim} gives that,
for $i\geq 2$,
\be\label{62-bis}
 \Pr(A_\mathds{1}) \E(Y_i|A_\mathds{1}^+) =2 \left(\frac{c}{3}\right)^{i-1} |I_1| \left(1-\frac{1}{|I_{i}|}\right).
\ee 
Finally, using equations \eqref{eq:case_i=1}, \eqref{eq:formula_P1E1} and \eqref{eq:induction_claim} we derive, for fixed $i$, the bound \eqref{62-again}:
\[
\Pr(A_\mathds{1}) \E(Y_i|A_\mathds{1}^+) \leq 2\left(\frac{c}{3}\right)^{i-1} |I_1|.
\]
\epr

\begin{example} \label{ex:example_lemma}
    In this example we consider a concrete landscape to clarify the proof of Lemma \ref{lemma:contributions}. This landscape and a specific trajectory for the embedded random walk are depicted below.

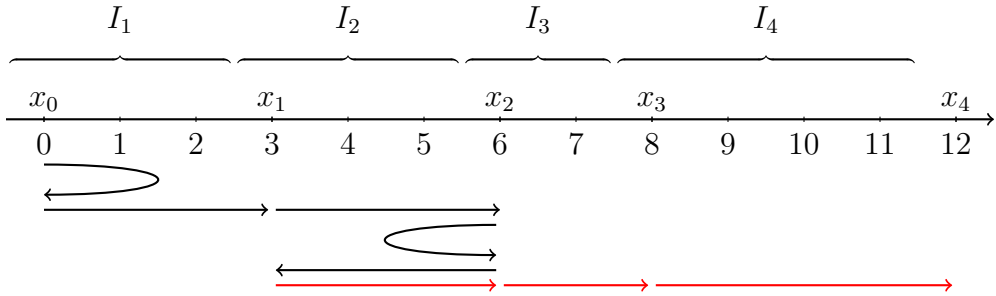
\begin{center}
\begin{tikzpicture}
\draw[thick,->] (-0.5,0) -- (12.5,0) ;
\foreach \x in {0,1,2,3,4,5,6,7,8,9,10,11,12}
   \draw (\x cm,1pt) -- (\x cm,-1pt) node[anchor=north] {$\x$};
   
\draw (0 cm,1pt) -- (0 cm,-1pt) node[anchor=south] {$x_0$};
\draw (3 cm,1pt) -- (3 cm,-1pt) node[anchor=south] {$x_1$};
\draw (6 cm,1pt) -- (6 cm,-1pt) node[anchor=south] {$x_2$};
\draw (8 cm,1pt) -- (8 cm,-1pt) node[anchor=south] {$x_3$};
\draw (12 cm,1pt) -- (12 cm,-1pt) node[anchor=south] {$x_4$};

\draw [thick, decorate, decoration = {calligraphic brace}] (-0.45,0.75) --  (2.45,0.75) node[pos=0.5,above = 7pt,black]{$I_1$};
\draw [thick, decorate, decoration = {calligraphic brace}] (2.55,0.75) --  (5.45,0.75) node[pos=0.5,above = 7pt,black]{$I_2$};
\draw [thick, decorate, decoration = {calligraphic brace}] (5.55,0.75) --  (7.45,0.75) node[pos=0.5,above = 7pt,black]{$I_3$};
\draw [thick, decorate, decoration = {calligraphic brace}] (7.55,0.75) --  (11.45,0.75) node[pos=0.5,above = 7pt,black]{$I_4$};


\draw[->,thick] (0,-0.6) .. controls (2,-0.6) and (2,-1) .. (0,-1);
\draw[->,thick] (0,-1.2) -- (2.95,-1.2);
\draw[->,thick] (3.05,-1.2) -- (6,-1.2);
\draw[->,thick] (5.95,-1.4) .. controls (4,-1.4) and (4,-1.8) .. (5.95,-1.8);
\draw[->,thick] (5.95,-2) -- (3.05,-2);
\draw[->,thick,color = red] (3.05,-2.2) -- (5.95,-2.2);
\draw[->,thick,color = red] (6.05,-2.2) -- (7.95,-2.2);
\draw[->,thick,color = red] (8.05,-2.2) -- (11.95,-2.2);
\end{tikzpicture}
\end{center}

The embedded random walk follows the sequence 

\be
\kappa = (0,1,1,0,-1,\color{red}1,1,1\color{black}) \in \mathcal{K} \subset \{-1,0,1\}^{8}.
\ee
The tail of ones at the end of $\kappa$ and the corresponding transitions of the embedded walk are indicated in red. Here $i=9$, $s=6$ and $\Theta = \sum_{j=1}^{s-1} \kappa_j =  1$. Note that we have $s<i-2$. Equation \eqref{eq:PA_kappa} becomes

\be \label{eq:PA_kappa_example}
\Pr(A_\kappa) = \Pr(A_{[\kappa]_1^5}) \cdot \color{red} \left(\frac{2}{3}\right)\left( \frac{1}{2|I_2|}\right) \cdot \left(\frac{2}{3}\right) \left(\frac{1}{2|I_3|}\right) \cdot \left(\frac{2}{3} \right)\left(\frac{1}{2|I_4|} \right)\color{black}.
\ee
Here, the factors corresponding to the tail of ones are indicated in red. We have $f = \Theta +i-s+1 = 5$. Equation \eqref{eq:EY_kappa} becomes

\be \label{eq:EY_kappa_example}
\E(Y_9|A_{\kappa}^+) =  |I_5|-1.
\ee

\subsubsection*{Replacing entry number $s$ by $\mathbf{\sigma = 0}$}

We now consider the sequence

\be
\kappa^{6,0} = (0,1,1,0,-1, \color{blue} 0 ,\color{red} 1,1 \color{black}).
\ee
The corresponding path of the embedded random walk is depicted below.

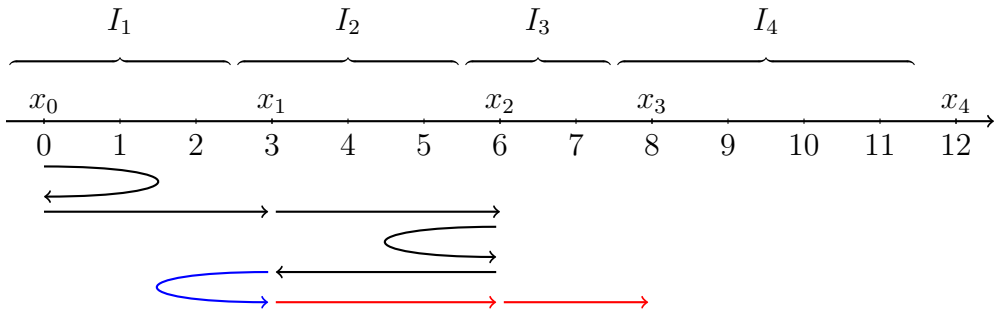
\begin{center}
\begin{tikzpicture}
\draw[thick,->] (-0.5,0) -- (12.5,0) ;
\foreach \x in {0,1,2,3,4,5,6,7,8,9,10,11,12}
   \draw (\x cm,1pt) -- (\x cm,-1pt) node[anchor=north] {$\x$};
   
\draw (0 cm,1pt) -- (0 cm,-1pt) node[anchor=south] {$x_0$};
\draw (3 cm,1pt) -- (3 cm,-1pt) node[anchor=south] {$x_1$};
\draw (6 cm,1pt) -- (6 cm,-1pt) node[anchor=south] {$x_2$};
\draw (8 cm,1pt) -- (8 cm,-1pt) node[anchor=south] {$x_3$};
\draw (12 cm,1pt) -- (12 cm,-1pt) node[anchor=south] {$x_4$};

\draw [thick, decorate, decoration = {calligraphic brace}] (-0.45,0.75) --  (2.45,0.75) node[pos=0.5,above = 7pt,black]{$I_1$};
\draw [thick, decorate, decoration = {calligraphic brace}] (2.55,0.75) --  (5.45,0.75) node[pos=0.5,above = 7pt,black]{$I_2$};
\draw [thick, decorate, decoration = {calligraphic brace}] (5.55,0.75) --  (7.45,0.75) node[pos=0.5,above = 7pt,black]{$I_3$};
\draw [thick, decorate, decoration = {calligraphic brace}] (7.55,0.75) --  (11.45,0.75) node[pos=0.5,above = 7pt,black]{$I_4$};


\draw[->,thick] (0,-0.6) .. controls (2,-0.6) and (2,-1) .. (0,-1);
\draw[->,thick] (0,-1.2) -- (2.95,-1.2);
\draw[->,thick] (3.05,-1.2) -- (6,-1.2);
\draw[->,thick] (5.95,-1.4) .. controls (4,-1.4) and (4,-1.8) .. (5.95,-1.8);
\draw[->,thick] (5.95,-2) -- (3.05,-2);

\draw[->,thick, color = blue] (2.95,-2) .. controls (1,-2) and (1,-2.4) .. (2.95,-2.4);

\draw[->,thick,color = red] (3.05,-2.4) -- (5.95,-2.4);
\draw[->,thick,color = red] (6.05,-2.4) -- (7.95,-2.4);
\end{tikzpicture}
\end{center}

Notice that the path according to $\kappa^{6,0}$ ends at $x_3$ rather than $x_4$, as it did for the unperturbed sequence $\kappa$. Instead of the transition from $x_3$ to $x_4$, there is the transition from $x_1$ back to $x_1$. We hence have

\be \label{eq:PA_kappa_0_ex}
\Pr(A_{\kappa^{6,0}}) = \Pr(A_{[\kappa]_1^5}) \cdot \color{blue} \left(\frac{2}{3}\right)\left(1-\frac{1}{2|I_1|}- \frac{1}{2|I_2|}\right) \cdot \color{red} \left(\frac{2}{3}\right)\left( \frac{1}{2|I_2|}\right) \cdot \left(\frac{2}{3} \right)\left(\frac{1}{2|I_3|} \right)\color{black}.
\ee
Here, the factor corresponding to entry $s=6$ in $\kappa^{s,0}$ is indicated in blue, whereas the factors corresponding to the tail of ones are indicated in red. This equation corresponds to \eqref{eq:PA_sig=0}. The interval  $[x_3,x_4)$ has length $|I_4|$,  hence

\be \label{eq:EY_sig=0_ex}
\E(Y_9|A^+_{\kappa^{6,0}}) = |I_4| -1= |I_{f-1}| -1.
\ee
This equation corresponds to \eqref{eq:EY_sig=0}.\\

\subsubsection*{Replacing entry number $s$ by $\mathbf{\sigma = -1}$}

For the sequence 

\be
\kappa^{6,-1} = (0,1,1,0,-1, \color{blue} -1 ,\color{red} 1,1 \color{black}),
\ee
the corresponding path of the embedded walk is depicted below.

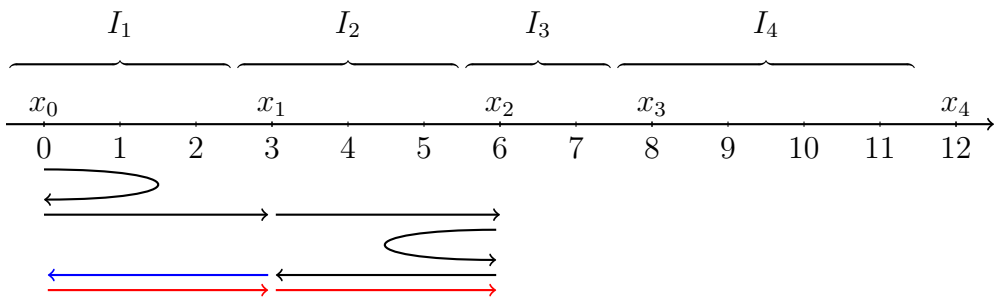
\begin{center}
\begin{tikzpicture}
\draw[thick,->] (-0.5,0) -- (12.5,0) ;
\foreach \x in {0,1,2,3,4,5,6,7,8,9,10,11,12}
   \draw (\x cm,1pt) -- (\x cm,-1pt) node[anchor=north] {$\x$};
   
\draw (0 cm,1pt) -- (0 cm,-1pt) node[anchor=south] {$x_0$};
\draw (3 cm,1pt) -- (3 cm,-1pt) node[anchor=south] {$x_1$};
\draw (6 cm,1pt) -- (6 cm,-1pt) node[anchor=south] {$x_2$};
\draw (8 cm,1pt) -- (8 cm,-1pt) node[anchor=south] {$x_3$};
\draw (12 cm,1pt) -- (12 cm,-1pt) node[anchor=south] {$x_4$};

\draw [thick, decorate, decoration = {calligraphic brace}] (-0.45,0.75) --  (2.45,0.75) node[pos=0.5,above = 7pt,black]{$I_1$};
\draw [thick, decorate, decoration = {calligraphic brace}] (2.55,0.75) --  (5.45,0.75) node[pos=0.5,above = 7pt,black]{$I_2$};
\draw [thick, decorate, decoration = {calligraphic brace}] (5.55,0.75) --  (7.45,0.75) node[pos=0.5,above = 7pt,black]{$I_3$};
\draw [thick, decorate, decoration = {calligraphic brace}] (7.55,0.75) --  (11.45,0.75) node[pos=0.5,above = 7pt,black]{$I_4$};


\draw[->,thick] (0,-0.6) .. controls (2,-0.6) and (2,-1) .. (0,-1);
\draw[->,thick] (0,-1.2) -- (2.95,-1.2);
\draw[->,thick] (3.05,-1.2) -- (6,-1.2);
\draw[->,thick] (5.95,-1.4) .. controls (4,-1.4) and (4,-1.8) .. (5.95,-1.8);
\draw[->,thick] (5.95,-2) -- (3.05,-2);

\draw[->,thick, color = blue] (2.95,-2) -- (0.05,-2);

\draw[->,thick, color = red] (0.05,-2.2) -- (2.95,-2.2);
\draw[->,thick,color = red] (3.05,-2.2) -- (5.95,-2.2);

\end{tikzpicture}
\end{center}

Notice that the path according to $\kappa^{6,-1}$ ends at $x_2$ rather then at $x_4$, as it did for the unperturbed sequence $\kappa$. Instead of the transitions from $x_2$ to $x_3$ and $x_3$ to $x_4$, we have the transition from $x_1$ to $x_0$ (corresponding to entry $s=6$ of $\kappa^{s,-1}$) and the transition from $x_0$ to $x_1$ (corresponding to the first one in the tail of $\kappa^{s,-1}$). Equation \eqref{eq:PA_sig=-1} becomes

\be \label{eq:PA_kappa_-1_ex}
\Pr(A_{\kappa^{6,-1}}) = \Pr(A_{[\kappa]_1^5}) \cdot \color{blue} \left(\frac{2}{3}\right)\left( \frac{1}{2|I_1|} \right)\cdot \color{red} \left(\frac{2}{3} \right)\left(\frac{1}{|I_1|} \right)\cdot\left( \frac{2}{3}\right)\left( \frac{1}{2|I_2|} \right)\color{black}.
\ee
Again, the factor corresponding to entry $s=6$ in $\kappa^{s,-1}$ is indicated in blue and those corresponding to the tail of ones are indicated in red. The interval  $[x_2,x_3)$  has length $|I_3|$ hence 

\be \label{eq:EY_sig=-1_ex}
\E(Y_9|A^+_{\kappa^{6,0}}) = |I_3|-1 = |I_{f-2}|-1.
\ee
This equation corresponds to \eqref{eq:EY_sig=-1}.

\end{example}

\subsection{Finalizing the proof of Theorem \ref{thm:criticality}}

Using Lemma \ref{lemma:contributions}, we now prove Theorem \ref{thm:criticality}. 

Recall \eqref{Etau-2}. We first prove that
\be \label{eq:ineq_+}
\E\left(   Y_i | A_\kappa  \right) \leq \E\left(   Y_i | A_\kappa^+  \right).
\ee
This is true because the interval lengths $|I_j|$ are increasing in $j$. This implies that if the embedded walk follows $\kappa$ and ends up at some trap $x_{f-1}$, the distance to $x_{f-2}$ is always smaller than the distance to $x_f$. As was explained in the proof of Lemma \ref{lemma:contributions} (see Equation \eqref{eq:EY_kappa} and below) the expectation of $Y_i$ conditioned on $S_{\mathcal{T}_{i-1} + 1}$ is equal to the distance between the first trap to the left and the first trap to the right of $S_{\mathcal{T}_{i-1} + 1}$. Because the distance $|x_f - x_{f-1}| = |I_f|$ is bigger than $|x_{f-1} - x_{f-2}| = |I_{f-1}|$, the expectation of $Y_i$ becomes larger if we condition on $A^+_\kappa$ rather than on $A_\kappa$, which explains \eqref{eq:ineq_+}.\\

\subsubsection{Proof of item 1 in Theorem \ref{thm:criticality}}

From \eqref{Etau-2} and \eqref{eq:ineq_+} we obtain the following inequalities
\begin{align}\label{54}
    \E(\tau) &\leq\sum_{i=1}^\infty \sum_{\kappa \in \mathcal{K}}\Pr(A_\kappa) \E\left(   Y_i | A_\kappa^+  \right) \\
    &\leq \sum_{i=1}^{\infty} \sum_{\kappa \in \{-1,0,1\}^{i-1}}\Pr(A_\kappa) \E\left(   Y_i | A_\kappa^+  \right) \nonumber \\
    &\leq \sum_{i=1}^\infty \sum_{l=0}^{i-1} \sum_{j=0}^{l} \binom{i-1}{l} \binom{l}{j} \left(\frac{2}{c}\right)^{l-j} \left(\frac{2}{c^2|I_1|^2} \right) ^{j} \cdot \Pr(A_\mathds{1}) \E(Y_i|A^+_\mathds{1}).\nonumber
\end{align}
Lemma \ref{lemma:contributions} is used in the last inequality, where $l$ should be interpreted as number of entries in $\kappa$ different from one and $j$ should be interpreted as the number of entries equal to $-1$ (i.e. $J$ in Lemma \ref{lemma:contributions}). We then have that $l-j$ is the number of entries equal to zero (i.e. $L$ in the Lemma \ref{lemma:contributions}). For the second inequality we sum over all  $\kappa \in \{-1,0,1\}^{i-1}$, rather than only over 
$\kappa$ in $\mathscr{K}$. This is an upper bound because we only add extra positive terms. We can now use the binomial theorem to see that
\begin{align}\label{55}
    \sum_{l=0}^{i-1} \sum_{j=0}^l \binom{i-1}{l} \binom{l}{j} \Big(\frac{2}{c}\Big)^{l-j} \Big(\frac{2}{c^2 |I_1|^2}\Big)^j &= \sum_{l=0}^{i-1}\binom{i-1}{l}  \left(\frac{2}{c}+ \frac{2}{c^2|I_1|^2}\right)^l\\
    &= \left(1+ \frac{2}{c} +\frac{2}{c^2 |I_1|^2}\right)^{i-1}\nonumber
\end{align}
By \eqref{54} and \eqref{55} we obtain,
\be
\label{eq:ubd_tau}
    \E(\tau) \leq \sum_{i=1}^\infty \left(1+ \frac{2}{c} +\frac{2}{c^2 |I_1|^2}\right)^{i-1} \cdot\Pr(A_\mathds{1}) \E(Y_i|A_\mathds{1}^+).
\ee
We substitute the bound \eqref{62-again} in \eqref{eq:ubd_tau} and obtain
\[
\E(\tau) \leq 2 |I_1| \sum_{i=1}^\infty \left(1+ \frac{2}{c} +\frac{2}{c^2 |I_1|^2}\right)^{i-1} \cdot \left(\frac{c}{3}\right)^{i-1}.
\]
The right hand side converges whenever 
\be\label{eq:condconvitem1}
 \left(1+ \frac{2}{c} +\frac{2}{c^2 |I_1|^2}\right)\cdot \left(\frac{c}{3}\right) < 1.
\ee
or, equivalently $c^2 |I_1|^2 - c|I_1|^2 +2 < 0$. We conclude that $\E(\tau)< \infty$ if $c$ satifies the following bounds
\[
\frac{|I_1| - \sqrt{|I_1|^2-8} }{2|I_1|}  < c < \frac{|I_1|+ \sqrt{|I_1|^2-8}}{2|I_1|}.
\]
Notice that the lower bound for $c$ is trivially true because of the assumptions $c > |I_1|^{-1}$ and $|I_1| \geq 3$. Indeed, for $|I_1|\geq 3$
\[
\frac{|I_1| - \sqrt{|I_1|^2-8} }{2|I_1|} \leq \frac{1}{|I_1|} < c.
\]
Therefore $c$ has to satisfy the bound \eqref{eq:c<}.
\qed
\subsubsection{Proof of item 2 in Theorem \ref{thm:criticality}}
As before, we have  (recall \eqref{Etau-2})
\begin{align}
    \E(\tau) 
    &= \sum_{i=1}^\infty \sum_{\kappa \in \mathcal{K}}\Pr(A_\kappa) \E\left(   Y_i | A_\kappa  \right). \nonumber
\end{align}
We only consider contributions from $\kappa \in \{0,1\}^{i-1}=: \mathscr{K}_0 \subset \mathscr{K}$ to obtain a lower bound for $\E(\tau)$. That is, we ignore the contributions of $\kappa$ with entries equal to minus one.

\begin{align} \label{eq:lbd_tau}
    \E(\tau) &\geq \sum_{i=1}^\infty \sum_{\kappa \in \{0,1\}^{i-1}}\Pr(A_\kappa) \E\left(   Y_i | A_\kappa  \right)\\
    &\geq \sum_{i=1}^\infty \sum_{\kappa \in \{0,1\}^{i-1}}\Pr(A_\kappa) \Pr(A^+_{\kappa}| A_\kappa) \E\left(   Y_i | A^+_\kappa  \right)\nonumber\\
    &\geq \sum_{i=1}^\infty \sum_{\kappa \in \{0,1\}^{i-1}} \frac{1}{3}\Pr(A_\kappa) \E\left(   Y_i | A^+_\kappa  \right) \nonumber\\
    &\geq \sum_{i=1}^\infty \sum_{l=0}^{i-1}\binom{i-1}{l} \frac{1}{6} \Big( \frac{2}{c}\Big(1- \frac{1}{|I_1|}\Big)^2\Big)^l \Pr ( A_{\mathds{1}}) \E(Y_i|A_\mathds{1}^+) \nonumber
\end{align}
For the second inequality we used $A_{\kappa}\supset A^+_{\kappa}$ and therefore
\[
\E(Y_i .\mathbb{I}_{A_{\kappa}})\geq \E(Y_i .\mathbb{I}_{A^+_{\kappa}})
\]
For the third inequality we use that the probability of the event $A^+_{\kappa}$ is greater or equal to $1/3$ the probability of the event $A_{\kappa}$. This is because in both events the walk follows $\kappa$, but in $A_\kappa^+$ the walk also has to survive and jump right afterwards. Survival and a jump to the right occur with probability $2/3$ if the walk is at $x_0=0$ and with probability $1/3$ if the walk is at any other trap. For the last inequality we used Lemma \ref{lemma:contributions}, where $l$  corresponds to the number of entries of $\kappa$ which are zero (i.e. $L$ in formula \eqref{eq:kappa_lbd} of Lemma \ref{lemma:contributions}). The binomial theorem gives

\be
\sum_{l=0}^{i-1} \binom{i-1}{l} \left(\frac{2}{c}\left(1-\frac{1}{|I_1|}\right)^2 \right)^l = \left(\frac{2}{c}\left(1-\frac{1}{|I_1|}\right)^2 + 1\right)^{i-1}.
\ee
Hence the right hand side of equation \eqref{eq:lbd_tau} becomes

\begin{align}\label{75} 
 \E(\tau) &\geq \sum_{i=1}^\infty \frac{1}{6}\left(\frac{2}{c}\left(1-\frac{1}{|I_1|}\right)^2 + 1\right)^{i-1} \Pr ( A_{\mathds{1}}) \E(Y_i|A_\mathds{1}^+)
\end{align}
We substitute 
in \eqref{75} the expressions we found for $\Pr ( A_{\mathds{1}}) \E(Y_i|A_\mathds{1}^+)$ in \eqref{62}. This gives

\begin{align}\label{76} 
 \E(\tau) &\geq \frac{1}{6}(|I_1|-1) + \sum_{i=2}^\infty \frac{1}{6}\left(\frac{2}{c}\left(1-\frac{1}{|I_1|}\right)^2 + 1\right)^{i-1} \cdot 2 \left( \frac{c}{3}\right)^{i-1}|I_1|\left(1-\frac{1}{|I_{1}|}\right).
\end{align}
We see that the right hand side of \eqref{76} diverges if 
\[
\left(\frac{2}{c}\left(1-\frac{1}{|I_1|}\right)^2 + 1\right) \cdot \left( \frac{c}{3}\right) \geq 1,
\]
or equivalently, if 
\begin{equation}\label{eq:equiv}
    c \geq 3 - 2\left( 1-\frac{1}{|I_1|}\right)^2 = 1+\frac{4}{|I_1|}-\frac{1}{|I_1|^2}.
\end{equation}
As was already explained in Remark \ref{rmk:possible_c}, the only possible values for $c$ are
\[
\frac{m}{|I_1|} \quad \text{for some } m \in \N.
\]
This is due to the restriction that all the intervals $I_i, i \geq 1$ have integer length. Since $1/|I_1|^2 < 1/|I_1|$, coming back to \eqref{eq:equiv}, there is no value for $m\in \N$ such that
\[
1 + \frac{4}{|I_1|} -\frac{1}{|I_1|^2}  \geq \frac{m}{|I_1|} > 1+\frac{3}{|I_1|}.
\]
We conclude that for all $c$ (which satisfy the integer restriction given $|I_1|$), $\E(\tau) = \infty$ if 
condition \eqref{eq:c>} is satisfied, that is
\[
c> 1+\frac{3}{|I_1|}.
\]
\qed

\section{Monotonicity} \label{sec:Monotonicity}

We start this section with an example of \emph{non-monotonicity} which follows directly from Theorem \ref{thm:main_result} (Example \ref{ex:nonintuitive}). We then continue with another class of counterexamples (in subsection 
\ref{subsec:counterexample}) showing that the expected survival time is not monotone as a function of the length of a single interval between two successive traps. We will then prove  Theorem \ref{thm:finiteness_tail} in subsection \ref{subsec:proof_finiteness_tail}, using the material derived in the previous subsection.  \\

\begin{example}\label{ex:nonintuitive}
In fact, Theorem 3.1 already
exhibits such non-monotonicity.

 Consider the setting of Theorem 3.1 with
the following values of $c$ and
$|I_1|$, reminding formula \eqref{eq:int_resriction}.
\begin{enumerate}
\item $c=2$, $|I_1|=1$ then
$|I_{n+1}|=  2^{2^n-1}$ and
$\E(\tau)=\infty$ (critical case)
\item $c=1/2$, $|I_1|=4$, then
$|I_{n+1}|= 4\cdot 2^{2^n-1}$ and
$\E(\tau)<\infty$ (subcritical case).
\end{enumerate}
This demonstrates that increasing the size of a single interval within a given trap landscape can, in fact, lead to a decrease in the expected survival time. Starting with a landscape where $c=2$ and $|I_1| = 1$, we can enlarge each interval, from left to right, by a factor of four to obtain the second landscape with $c=1/2$ and $|I_1| = 4$. If increasing the size of an interval always led to a longer survival time, then the expected survival time in the second landscape would necessarily be greater than in the first. However, since the first landscape is critical and the second is subcritical, this implies that the survival time of the random walk cannot, in general, increase monotonically with the length of a single interval.
\end{example}

To understand better this form of ``non-monotonicity'', we then  study the difference in expected survival time between two arbitrary landscapes (i.e. landscapes not necessarily satisfying the recursion in \eqref{eq:recursion}) and provide a class of examples where increasing a specific single interval by one yields a smaller expectation for the survival time. \\

We  begin with a precise definition of monotonicity. 
Consider an arbitrary landscape of traps $\om^{(1)}$. We will denote all the associated quantities introduced in Section \ref{sec:Random_walk_in_a_landscape_of_soft_traps} by adding a superscript $(1)$, e.g. the survival time is denoted $\tau^{(1)}$ and the intervals between traps are denoted $I^{(1)}_i, i \geq 1$. Fix an arbitrary  $k \in \{1, 2, \dots \}$.  We construct a second landscape, $\om^{(2)}$ by increasing the length of interval $I^{(1)}_k$ by one site and keeping all the other interval lengths $|I^{(1)}_l|, l \neq k$ unchanged, that is, $x_l^{(2)}=x_l^{(1)}$ for $l\leq k-1$, and 
$x_l^{(2)}=x_l^{(1)}+1$ for $l\geq k$; in other words: 

    \be \label{eq:om_1_2_monotonicity}
    \om^{(2)}(x)= 
    \begin{cases}
        \om^{(1)}(x) \quad &\text{if } x \leq x^{(1)}_{k-1}\\
        0 \quad &\text{if } x = x^{(1)}_{k-1} + 1 \\
        \om^{(1)}(x-1) \quad &\text{if } x > x^{(1)}_{k-1} + 1.
    \end{cases}
    \ee
We say that the expected survival time is \textit{monotone in} $|I_k|$ if

\be \label{eq:false_monotonicity_statement}
\E[\tau^{(1)}] \leq \E[\tau^{(2)}].
\ee

\noindent
\underline{Landscape $\om^{(1)}$:}

\begin{center}
\begin{tikzpicture}
\centering
\draw[thick,->] (-0.5,0) -- (12.5,0) ;
\foreach \x in {0,1,2,3,4,5,6,7,8,9,10,11,12}
   \draw (\x cm,1pt) -- (\x cm,-1pt);
   
\draw (0 cm,1pt) -- (0 cm,-1pt) node[anchor=south] {$x^{(1)}_{k-2}$};
\draw (3 cm,1pt) -- (3 cm,-1pt) node[anchor=south] {$x^{(1)}_{k-1}$};
\draw (7 cm,1pt) -- (7 cm,-1pt) node[anchor=south] {$x^{(1)}_{k}$};
\draw (11 cm,1pt) -- (11 cm,-1pt) node[anchor=south] {$x^{(1)}_{k+1}$};

\draw[blue, thick] (-0.5,-0.1) -- (3.25,-0.1);
\node[blue] at (1.5,0.75) {$\mathbb{S}_-$};
\draw[red, thick] (3.75,-0.1) -- (12.25,-0.1);
\node[red] at (9,0.75) {$\mathbb{S}_+$};

\draw [thick, decorate, decoration = {calligraphic brace,mirror}] (-0.5,-0.5) --  (2.5,-0.5) node[pos=0.5,below = 7pt,black]{$I^{(1)}_{k-1}$};
\draw [thick, decorate, decoration = {calligraphic brace,mirror}] (2.5,-0.5) --  (6.5,-0.5) node[pos=0.5,below = 7pt,black]{$I^{(1)}_k$};
\draw [thick, decorate, decoration = {calligraphic brace,mirror}] (6.5,-0.5) --  (10.5,-0.5) node[pos=0.5,below = 7pt,black]{$I^{(1)}_{k+1}$};
\end{tikzpicture}
\end{center}

\noindent
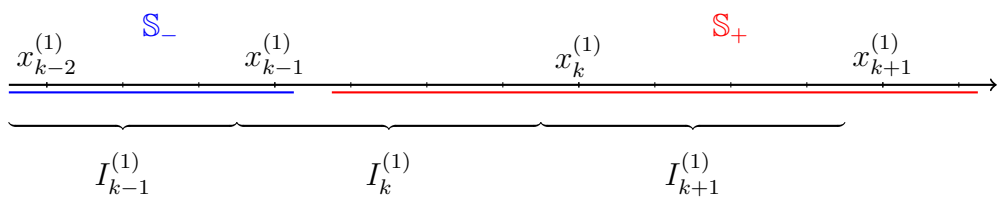
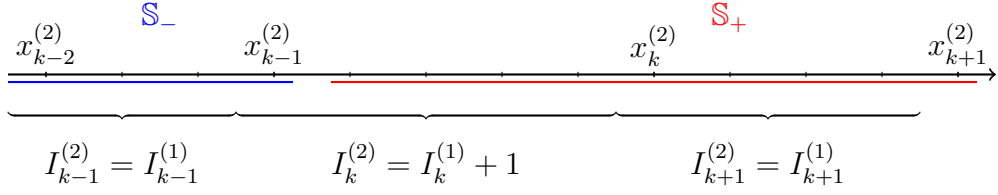
\underline{Landscape $\om^{(2)}$:}

\begin{center}
\begin{tikzpicture}
\centering
\draw[thick,->] (-0.5,0) -- (12.5,0) ;
\foreach \x in {0,1,2,3,4,5,6,7,8,9,10,11,12}
   \draw (\x cm,1pt) -- (\x cm,-1pt);
   
\draw (0 cm,1pt) -- (0 cm,-1pt) node[anchor=south] {$x^{(2)}_{k-2}$};
\draw (3 cm,1pt) -- (3 cm,-1pt) node[anchor=south] {$x^{(2)}_{k-1}$};
\draw (8 cm,1pt) -- (8 cm,-1pt) node[anchor=south] {$x^{(2)}_{k}$};
\draw (12 cm,1pt) -- (12 cm,-1pt) node[anchor=south] {$x^{(2)}_{k+1}$};

\draw[blue, thick] (-0.5,-0.1) -- (3.25,-0.1);
\node[blue] at (1.5,0.75) {$\mathbb{S}_-$};
\draw[red, thick] (3.75,-0.1) -- (12.25,-0.1);
\node[red] at (9,0.75) {$\mathbb{S}_+$};

\draw [thick, decorate, decoration = {calligraphic brace,mirror}] (-0.5,-0.5) --  (2.5,-0.5) node[pos=0.5,below = 7pt,black]{$I^{(2)}_{k-1} = I^{(1)}_{k-1}$};
\draw [thick, decorate, decoration = {calligraphic brace,mirror}] (2.5,-0.5) --  (7.5,-0.5) node[pos=0.5,below = 7pt,black]{$I^{(2)}_k = I^{(1)}_k +1 $};
\draw [thick, decorate, decoration = {calligraphic brace,mirror}] (7.5,-0.5) --  (11.5,-0.5) node[pos=0.5,below = 7pt,black]{$I^{(2)}_{k+1} = I^{(1)}_{k+1}$};
\end{tikzpicture}
\end{center}

For $\gamma \in \{1,2\}$ we introduce the random times at which the walker in $\om^{(\gamma)}$ makes a transition between the sets 
\beq\label{eq:S+}\mathbb{S}_+&:=&\{x^{(1)}_{k-1} +1, x^{(1)}_{k-1} +2, \dots \}\\ \label{S+}
\mathbb{S}_- &:=& \{0, 1, \dots,   x^{(1)}_{k-1}\}\eeq
 (since $x^{(1)}_{l}=x^{(2)}_{l}$ for $l\leq k-1$, we do not put superscripts $(1)$ and $(2)$ for $\mathbb{S}_+$ and $\mathbb{S}_-$ because they are the same for both landscapes): 

\begin{align}
    l^{(\gamma)}_0 &:= \inf\{t \geq 0: S^{(\gamma)}_t = x^{(1)}_{k-1}\}\\
    l_1^{(\gamma)} &:= \inf \left\{t \geq l^{(\gamma)}_0: S^{(\gamma)}_t = x^{(1)}_{k-1}+1 \right\}\nonumber\\
    l_2^{(\gamma)} &:= \inf \left\{t \geq l^{(\gamma)}_1: S^{(\gamma)}_t = x^{(1)}_{k-1} \right\}\nonumber\\
    l_3^{(\gamma)} &:= \inf \left\{t \geq l^{(\gamma)}_2: S^{(\gamma)}_t = x^{(1)}_{k-1}+1 \right\}\nonumber\\
    l_4^{(\gamma)} &:= \inf \left\{t \geq l^{(\gamma)}_3: S^{(\gamma)}_t = x^{(1)}_{k-1} \right\}\nonumber\\
    &\dots \nonumber
\end{align}
That is, we take $l_{-1}^{(\gamma)}:=0$ and for $i \geq0$ we define $l^{(\gamma)}_i$ as

\begin{align} \label{eq:l_i}
l^{(\gamma)}_i &:= \inf \Bigg \{t \geq l^{(\gamma)}_{i-1}: S^{(\gamma)}_t =
\begin{cases}
     x^{(1)}_{k-1} &\quad \text{if } S^{(\gamma)}_{l_{i-1}} = x^{(1)}_{k-1} + 1\\
     x^{(1)}_{k-1} + 1 &\quad \text{if } S^{(\gamma)}_{l_{i-1}} = x^{(1)}_{k-1}
\end{cases}
\Bigg\}.
\end{align}
Here we define the infimum of the empty set as $\infty$. This implies that $l^{(\gamma)}_{i+1} = \infty$ if the walker in 
 $\om^{(\gamma)}$ makes transitions between the sets $\mathbb{S}_+$ and $\mathbb{S}_-$ at most $i$ times before dying.

Next we write the survival time $\tau^{(\gamma)}$ as a sum of times spent in the sets  $\mathbb{S}_+$ and  $\mathbb{S}_-$. More precisely, we define 
\be\label{eq:Tgamman}
T^{(\gamma)}_n := \tau^{(\gamma)} \wedge l^{(\gamma)}_{n} - \tau^{(\gamma)} \wedge l^{(\gamma)}_{n-1}
\ee
for $n \geq 0$. Because $l^{(\gamma)}_{-1} = 0$, we have
\be\label{eq:Tgamma0}
T^{(\gamma)}_0 := \tau^{(\gamma)} \wedge l^{(\gamma)}_0.
\ee
Then we have
\be\label{eq:tau-with-Tgamma}
\tau^{(\gamma)} = \sum_{n=0}^{\infty
} T^{(\gamma)}_n.
\ee
Suppose that $n = 2m,$ $m \in \N\setminus\{0\}$ is even. Then, one can think of $T^{(\gamma)}_n$ as the time spent in the set $\mathbb{S}_+$ after the walker entered $\mathbb{S}_+$ for the $m$-th time. Similarly, for $n = 2m+1$ $m \in \N$, one can think of $T^{(\gamma)}_n$ as the time spent in the set $\mathbb{S}_-$ after the walker entered $\mathbb{S}_-$ for the $m$-th time. The random variable $T^{(\gamma)}_0$ is the time it takes to the walker to reach $x^{(1)}_{k-1} = x^{(2)}_{k-1}$ for the first time.

\begin{example}\label{ex:first}
Fix $\gamma \in \{1,2\}$. Depicted below is an example of a landscape $\om^{(\gamma)}$ and a trajectory of the random walk $S^{(\gamma)}$. The trajectory is such that the walker gets trapped at $x_1^{(\gamma)}$ at time $\tau^{(\gamma)} = 28$. 

In this example we take $k=3$ and we indicate the corresponding random times $l_i^{(\gamma)}, i \in \{0,1,2,3,4\}$, as well as the times $T^{(\gamma)}_n, n\geq 1$. The set $\mathbb{S}_-$ and the times at which the walker enters $\mathbb{S}_-$ are marked in blue. Similarly, the set $\mathbb{S}_+$ and the times when the random walker enters $\mathbb{S}_+$ are marked in red.
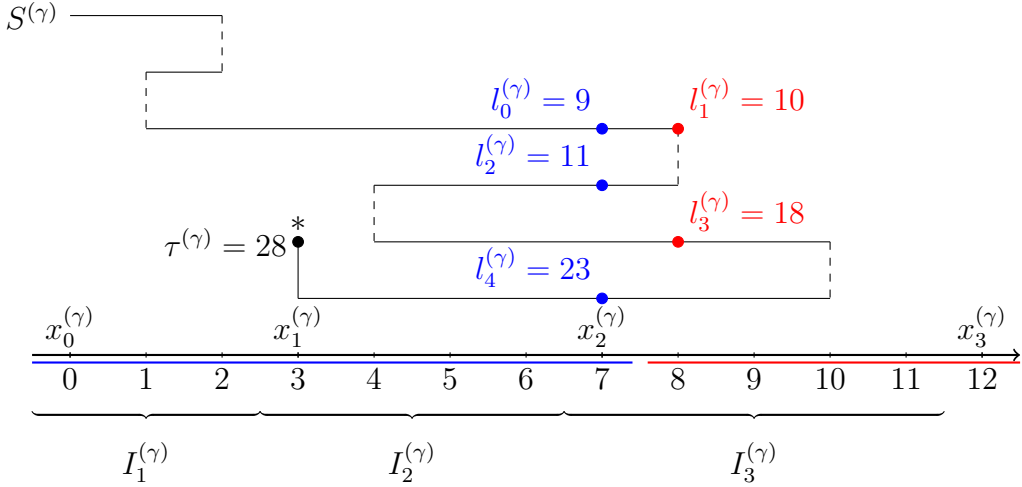
\begin{center}
\begin{tikzpicture}

\path[-]
    node at (-0.5cm,4.5cm){$S^{(\gamma)}$}
    (0cm,4.5cm) edge(2cm, 4.5cm)
    (2cm, 4.5cm) edge[dashed](2cm,3.75cm)
    
    (2cm,3.75cm) edge(1cm, 3.75cm)
    (1cm, 3.75cm) edge[dashed](1cm,3cm)
    
    (1cm, 3cm) edge(8cm, 3cm)
    (8cm, 3cm) edge[dashed](8cm,2.25cm)
    
    (8cm, 2.25cm) edge(4cm, 2.25cm)
    (4cm, 2.25cm) edge[dashed](4cm,1.5cm)
    
    (4cm, 1.5cm) edge(10cm, 1.5cm)
    (10cm, 1.5cm) edge[dashed](10cm,0.75cm)
    
    (10cm, 0.75cm) edge (3cm, 0.75cm)
    
    (3cm, 0.75cm) edge (3cm, 1.5cm);

    \filldraw[blue] (7,3) circle (2pt) node[anchor=south east]{$l^{(\gamma)}_0 = 9$};
    \filldraw[red] (8,3) circle (2pt) node[anchor= south west]{$l^{(\gamma)}_1 = 10$};
    \filldraw[blue] (7,2.25) circle (2pt) node[anchor=south east]{$l^{(\gamma)}_2 = 11$};
    \filldraw[red] (8,1.5) circle (2pt) node[anchor= south west]{$l^{(\gamma)}_3 = 18$};
    \filldraw[blue] (7,0.75) circle (2pt) node[anchor=south east]{$l^{(\gamma)}_4 = 23$};
    \filldraw[black] (3,1.5) circle (2pt) node[anchor= east]{$\tau^{(\gamma)} = 28$} node[anchor=south]{$\ast$};

\draw[thick,->] (-0.5,0) -- (12.5,0) ;
\foreach \x in {0,1,2,3,4,5,6,7,8,9,10,11,12}
   \draw (\x cm,1pt) -- (\x cm,-1pt) node[anchor=north] {$\x$};
   
\draw (0 cm,1pt) -- (0 cm,-1pt) node[anchor=south] {$x^{(\gamma)}_0$};
\draw (3 cm,1pt) -- (3 cm,-1pt) node[anchor=south] {$x^{(\gamma)}_1$};
\draw (7 cm,1pt) -- (7 cm,-1pt) node[anchor=south] {$x^{(\gamma)}_2$};
\draw (12 cm,1pt) -- (12 cm,-1pt) node[anchor=south] {$x^{(\gamma)}_3$};

\draw [thick, decorate, decoration = {calligraphic brace,mirror}] (-0.5,-0.75) --  (2.5,-0.75) node[pos=0.5,below = 7pt,black]{$I^{(\gamma)}_1$};

\draw [thick, decorate, decoration = {calligraphic brace,mirror}] (2.5,-0.75) --  (6.5,-0.75) node[pos=0.5,below = 7pt,black]{$I^{(\gamma)}_2$};
\draw [thick, decorate, decoration = {calligraphic brace,mirror}] (6.5,-0.75) --  (11.5,-0.75) node[pos=0.5,below = 7pt,black]{$I^{(\gamma)}_3$};
\draw[thick, blue] (-0.5,-0.1) -- (7.4,-0.1) ;
\draw[thick, red] (7.6,-0.1) -- (12.5,-0.1) ;

\end{tikzpicture}
\end{center}

For the times $T^{(\gamma)}_n, n\geq 1$, we find
\begin{align}\label{eq:T-ex}
    &T^{(\gamma)}_0 = \tau^{(\gamma)} \wedge l^{(\gamma)}_0 = 9\\
    &T^{(\gamma)}_1 = \tau^{(\gamma)} \wedge l^{(\gamma)}_{1} - \tau^{(\gamma)} \wedge l^{(\gamma)}_{0} = 10 -9 =1 \nonumber \\
    &T^{(\gamma)}_2 = \tau^{(\gamma)} \wedge l^{(\gamma)}_{2} - \tau^{(\gamma)} \wedge l^{(\gamma)}_{1} = 11 - 10 =1 \nonumber \\
    &T^{(\gamma)}_3 = \tau^{(\gamma)} \wedge l^{(\gamma)}_{3} - \tau^{(\gamma)} \wedge l^{(\gamma)}_{2} = 18 - 11 = 7 \nonumber \\
    &T^{(\gamma)}_4 = \tau^{(\gamma)} \wedge l^{(\gamma)}_{4} - \tau^{(\gamma)} \wedge l^{(\gamma)}_{3} = 23 - 18 = 5 \nonumber \\
    &T^{(\gamma)}_5 = \tau^{(\gamma)} \wedge l^{(\gamma)}_{5} - \tau^{(\gamma)} \wedge l^{(\gamma)}_{4} = \infty \wedge 28 - 23 = 5 \nonumber
\end{align} 

and
\be\label{eq:Tn-ex}
 T^{(\gamma)}_n = 0 \qquad \text{for }n > 5.
\ee
We indeed have
\be\label{eq:tau-ex}
\tau^{(\gamma)} = T^{(\gamma)}_0 + T^{(\gamma)}_1 + \dots +T^{(\gamma)}_5 = 28.
\ee
\end{example}

Notice that the distribution of $l^{(\gamma)}_i$ given $l^{(\gamma)}_{i-1},..., l^{(\gamma)}_{0}$ only depends on  $l^{(\gamma)}_{i-1}$ due to the Markov property of the process $\{S^{(\gamma)}_n, n \in \N \}$. Similarly the distribution of $\tau^{(\gamma)}$ given $l^{(\gamma)}_{i-1},..., l^{(\gamma)}_{0}$ only depends on $l^{(\gamma)}_{i-1}$. As a consequence, the quantities 
\be\label{eq:E+andE-}
    \caE^{(\gamma)}_+ := \E[T^{(\gamma)}_{2m} | l^{(\gamma)}_{2m-1} < \infty ] \quad \text{and} \quad  \caE^{(\gamma)}_- := \E[T^{(\gamma)}_{2m+1} | l^{(\gamma)}_{2m} < \infty]
\ee
as well as
\be\label{eq:P+andP-}
\caP^{(\gamma)}_+:= \Pr(l^{(\gamma)}_{2m} < \infty| l^{(\gamma)}_{2m-1} < \infty ) \quad \text{and} \quad \caP^{(\gamma)}_-:= \Pr(l^{(\gamma)}_{2m+1} < \infty| l^{(\gamma)}_{2m} < \infty ),
\ee
are all independent of $m$ for $m>0$. We additionally define
\be\label{eq:E0andP0}
\caE^{(\gamma)}_0 := \E[T^{(\gamma)}_0] \quad \text{and} \quad \caP^{(\gamma)}_0 := \Pr(l^{(\gamma)}_0 <\infty)
\ee
The number $\caE^{(\gamma)}_+$ should be interpreted as the expected time it takes for the walker,  after having first entered the set $\mathbb{S}_+$,  to either leave this set $\mathbb{S}_+$ and enter 
$\mathbb{S}_-$, or get trapped before leaving $\mathbb{S}_+$.  Similarly, the number $\caE^{(\gamma)}_-$ should be interpreted as the expected time it takes for the walker, after having first entered the set $\mathbb{S}_-$,  to either leave this set $\mathbb{S}_-$ and enter 
$\mathbb{S}_+$, or get trapped before leaving $\mathbb{S}_-$. 
The probabilities $\caP^{(\gamma)}_+$ and $\caP^{(\gamma)}_-$ are the probabilities that the random walker exits the sets $\mathbb{S}_+$ and $\mathbb{S}_-$ respectively after entering them. In Lemma \ref{lemma:expression_suvival_curly} the expected survival time corresponding to $\gamma$ is stated as a function of the quantities $\mathcal{E}^{(\gamma)}_+$, $\mathcal{E}^{(\gamma)}_0$, $\mathcal{E}^{(\gamma)}_-$ and $\mathcal{P}^{(\gamma)}_+$, $\mathcal{P}^{(\gamma)}_0$, $\mathcal{P}^{(\gamma)}_-$.

\bl \label{lemma:expression_suvival_curly}
Let $\gamma \in \{1,2\}$. The expectation of the survival time $\tau^{(\gamma)}$ is given by 
\be\label{eq:expression_suvival_curly}
\E[\tau^{(\gamma)}] = \caE^{(\gamma)}_0 +  \caP^{(\gamma)}_0 \Big[ \caE^{(\gamma)}_-  \frac{1}{1-\caP^{(\gamma)}_+ \caP^{(\gamma)}_-} + \caE^{(\gamma)}_+ \frac{\caP^{(\gamma)}_-}{1-\caP^{(\gamma)}_+ \caP^{(\gamma)}_-} \Big].
\ee
\el

\bpr
We leave out the superscript $(\gamma)$ since the argument holds both for $\gamma = 1$ and $\gamma = 2$. We write the survival time $\tau$ as an infinite sum of time intervals (cf. \eqref{eq:Tgamma0}
and \eqref{eq:tau-with-Tgamma}), i.e. 
\be\label{eq:Etau-again}
\E[\tau] = \sum_{n=0}^{\infty} \E[T_n].
\ee
We can write the probability that a time $l_i$ is finite (i.e. the probability that the walker makes transitions at least $i$ times between $\mathbb{S}_+$ and $\mathbb{S}_-$ before moving to $\ast$) as (cf. \eqref{eq:P+andP-}, \eqref{eq:E0andP0})

\be\label{eq:lfinite}
\Pr(l_{2m+1} < \infty) = \caP_0 \caP_{-}(\caP_{-} \caP_{+})^m \quad \text{and} \quad \Pr(l_{2m} < \infty) =  \caP_0(\caP_{-} \caP_{+})^m.
\ee
This yields the following expression for the expected survival time (using also \eqref{eq:E+andE-}), 
\begin{align} \label{eq:expectation_tau}
\E[\tau] &= \sum_{n=0}^{\infty
} \Pr( l_{n-1} < \infty) \E[T_n |  l_{n-1} < \infty]\\\nonumber
&= \sum_{m=0}^{\infty
} \Pr( l_{2m-1} < \infty) \E[T_{2m} |  l_{2m-1} < \infty]
+ \sum_{m=0}^{\infty
} \Pr( l_{2m} < \infty) \E[T_{2m+1}|  l_{2m} < \infty]\\\nonumber
&= \caE_0 + \sum_{m=0}^{\infty
} \Pr( l_{2m+1} < \infty) \caE_+
+ \sum_{m=0}^{\infty} \Pr( l_{2m} < \infty) \caE_-\\\nonumber
&= \caE_0 +  \caP_0 \Big[ \caE_+ \caP_- \sum_{m=0}^\infty (\caP_+ \caP_-)^m + \caE_-  \sum_{m=0}^\infty (\caP_+ \caP_-)^m \Big]\\  \nonumber
&= \caE_0 +  \caP_0 \Big[ \caE_+ \frac{\caP_-}{1-\caP_+ \caP_-}+\caE_-  \frac{1}{1-\caP_+\caP_-}   \Big].\nonumber
\end{align}  
\epr

We want to compare the quantities $\E[\tau^{(1)}]$ and $\E[\tau^{(2)}]$ to see that monotonicity does not hold in general. Notice that the quantities $\caE^{(1)}_-$ and $\caE^{(2)}_-$  depend only on the positions of the first $k-1$ traps: $x^{(1)}_1, x^{(1)}_2, \dots, x^{(1)}_{k-1}$ and  $x^{(2)}_1, x^{(2)}_2, \dots, x^{(2)}_{k-1}$ respectively. Since these trap locations are the same in both landscapes, we have $\caE^{(1)}_- = \caE^{(2)}_-$. For the same reason, we have $\caE^{(1)}_0 = \caE^{(2)}_0$,$\caP^{(1)}_- = \caP^{(2)}_-$ and $\caP^{(1)}_0 = \caP^{(2)}_0$.The quantities $\caE^{(1)}_+$ and $\caE^{(2)}_+$ depend only on the positions of the traps with index $k$ or higher: $x^{(1)}_k, x^{(1)}_{k+1}, \dots$ and $x^{(2)}_k, x^{(2)}_{k+1}, \dots$ respectively. These traps are not the same so in general we  don not have $\caE^{(1)}_+ = \caE^{(2)}_+$. However, the locations of the traps with index higher or equal to $k$ only differ by one, i.e. $x^{(2)}_{k+i} = x^{(1)}_{k+i} + 1$. In other words, the whole trap landscape starting from $x^{(1)}_k$ in  $\om^{(1)}$ is shifted by one to the right in $\om^{(2)}$. We can use this to calculate $\caE^{(2)}_+$ as a function of $\caE^{(1)}_+$. Using a similar argument, we can also find $\caP^{(2)}_+$ as a function of $\caP^{(1)}_+$. This is what happens in the proof of Lemma \ref{Lemma:calculation_E_+^{(2)}}.

\bl \label{Lemma:calculation_E_+^{(2)}}
We have the following relations between $\caE^{(1)}_+$, $\caP^{(1)}_+$ and $\caE^{(2)}_+$, $\caP^{(2)}_+$,

\be\label{eq:calculation_E_+^{(2)}}
\caP^{(2)}_+ = \frac{1}{2- \caP^{(1)}_+} \qquad \text{and} \qquad \caE^{(2)}_+ = \frac{\caE^{(1)}_+ + 2}{2- \caP^{(1)}_+}.
\ee
\el

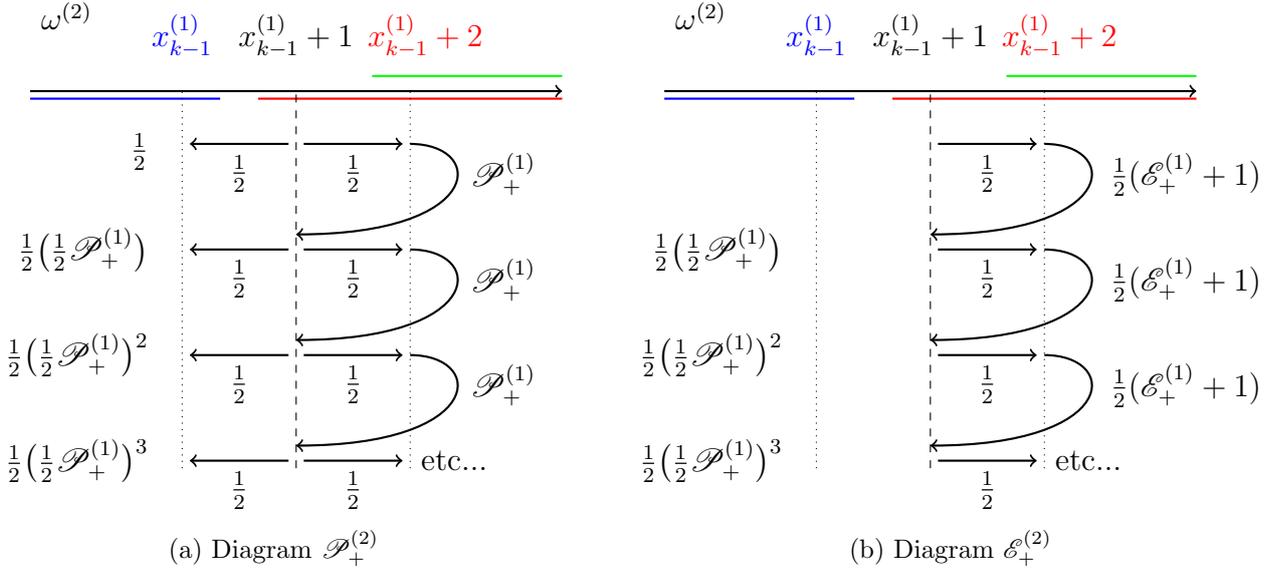
\bpr
\begin{center}
\begin{figure}[H]
\subfloat[Diagram $\caP^{(2)}_+$]
{
\label{subfig:P2_+}
\begin{tikzpicture}
    \node[anchor = west] at (-3.5,1) {$\om^{(2)}$};
    \node[anchor = south, black] at (0,0.3) {$x^{(1)}_{k-1} + 1$};
    \node[anchor = south, blue] at (-1.5,0.3) {$x^{(1)}_{k-1}$};
    \node[anchor = south,red] at (1.7,0.3) {$x^{(1)}_{k-1}+2$};
    \draw[thick,->] (-3.5,0) -- (3.5,0);
    
    \draw[blue, thick] (-3.5,-0.1) -- (-1,-0.1);
    \draw[red, thick] (-0.5,-0.1) -- (3.5,-0.1);
    \draw[green, thick] (1,0.2) -- (3.5,0.2);
    
    \draw[dashed] (0,-5) -- (0,0);
    \draw[dotted] (-1.5,-5) -- (-1.5,0);
    \draw[dotted] (1.5,-5) -- (1.5,0);
    
    \foreach \y in {-0.7,-2.1,-3.5} {
        \draw[->,thick] (1.5,\y) .. controls (2.5,\y) and (2.5,\y-1.2) .. (0,\y-1.2);
        \node[anchor = east] at (3.3,\y-0.4) {$\caP^{(1)}_+$};
    }

    \node[anchor = west] at (1.5,-4.9) {etc...};
    
    \foreach \y in {-0.7,-2.1,-3.5,-4.9} {
        \draw[->,thick] (-0.1,\y) -- (-1.4,\y);
        \draw[->,thick] (0.1,\y) -- (1.4,\y);
        \node[below] at (-0.75,\y) {$\frac{1}{2}$};
        \node[below] at (0.75,\y) {$\frac{1}{2}$};
    }

    \node[anchor = east] at (-1.8, -0.8) {$\frac{1}{2}$};
    \node[anchor = east] at (-1.8, -2.1) {$\frac{1}{2}\big(\frac{1}{2} \caP^{(1)}_+\big)$};
    \node[anchor = east] at (-1.8, -3.5) {$\frac{1}{2} \big(\frac{1}{2} \caP^{(1)}_+\big)^2$};
    \node[anchor = east] at (-1.8, -4.9) {$\frac{1}{2} \big(\frac{1}{2} \caP^{(1)}_+\big)^3$};
    
\end{tikzpicture}
}
\hspace{0.5cm}
\subfloat[Diagram $\caE^{(2)}_+$]
{\label{subfig:E2_+}
\begin{tikzpicture}
    \node[anchor = west] at (-3.5,1) {$\om^{(2)}$};
    \node[anchor = south, black] at (0,0.3) {$x^{(1)}_{k-1} + 1$};
    \node[anchor = south, blue] at (-1.5,0.3) {$x^{(1)}_{k-1}$};
    \node[anchor = south,red] at (1.7,0.3) {$x^{(1)}_{k-1}+2$};
    \draw[thick,->] (-3.5,0) -- (3.5,0);
    
    \draw[blue, thick] (-3.5,-0.1) -- (-1,-0.1);
    \draw[red, thick] (-0.5,-0.1) -- (3.5,-0.1);
    \draw[green, thick] (1,0.2) -- (3.5,0.2);
    
    \draw[dashed] (0,-5) -- (0,0);
    \draw[dotted] (-1.5,-5) -- (-1.5,0);
    \draw[dotted] (1.5,-5) -- (1.5,0);
    
    \foreach \y in {-0.7,-2.1,-3.5} {
        \draw[->,thick] (1.5,\y) .. controls (2.5,\y) and (2.5,\y-1.2) .. (0,\y-1.2);
        \node[anchor = east] at (4.5,\y-0.4) {$\frac{1}{2}(\caE^{(1)}_+ + 1)$};
    }

    \node[anchor = west] at (1.5,-4.9) {etc...};
    
    \foreach \y in {-0.7,-2.1,-3.5,-4.9} {
        \draw[->,thick] (0.1,\y) -- (1.4,\y);
        \node[below] at (0.75,\y) {$\frac{1}{2}$};
    }

    \node[anchor = east] at (-1.8, -2.1) {$\frac{1}{2}\big(\frac{1}{2} \caP^{(1)}_+\big)$};
    \node[anchor = east] at (-1.8, -3.5) {$\frac{1}{2} \big(\frac{1}{2} \caP^{(1)}_+\big)^2$};
    \node[anchor = east] at (-1.8, -4.9) {$\frac{1}{2} \big(\frac{1}{2} \caP^{(1)}_+\big)^3$};
    
\end{tikzpicture}
}
\end{figure}
\end{center}

In Figure \ref{subfig:P2_+} the trajectories contributing to $\caP^{(2)}_+$ are represented schematically. The area indicated in blue represents the trap landscape in $\mathbb{S}_-$, whereas the area indicated in red represents the trap landscape in $\mathbb{S}_+$. There is another area indicated in green. Here, the trap landscape looks exactly the same as the the landscape in $\om^{(1)}$ right of $x^{(1)}_{k-1}+1$. Therefore, the blue and the green areas are such that if one jumps into the green (respectively blue) area, the probability of leaving it before getting trapped is precisely $\caP^{(1)}_+$ (respectively $\caP^{(1)}_-$). Similarly, the expected time it takes to either get trapped or to leave the green (respectively blue) area  after entering is precisely $\caE^{(1)}_+$ (respectively $\caE^{(1)}_-$).

We prove the expression for $\caP^{(2)}_+$. Suppose that the walker just entered the set $\mathbb{S}_+ = \{x^{(1)}_{k-1}+1, x^{(1)}_{k-1}+2, x^{(1)}_{k-1}+3, \dots\}$, i.e. it is located at $x^{(1)}_{k-1}+1$. Then it can move in and out of the green area, $\{x^{(1)}_{k-1}+2, x^{(1)}_{k-1}+3, x^{(1)}_{k-1}+4, \dots\}$, an arbitrary number of times before moving out of $\mathbb{S}_+$. Each time the walker moves into the green area it does so with probability $1/2$. After entering, it leaves the green area again with probability $\caP^{(1)}_+$. After visiting the green area for $n \in \{0,1,2, \dots \}$ times, the walker might jump out of $\mathbb{S}_+$ with probability $1/2$. Such a trajectory hence occurs with probability $(1/2)(\caP^{(1)}_+/2)^n$. We find that the total probability of leaving $\mathbb{S}_+$ is given by

\be\label{eq:expressioncap}
\caP^{(2)}_+ = \frac{1}{2} \sum_{n=0}^{\infty} \Big(\frac{1}{2} \caP^{(1)}_+\Big)^n =\frac{1}{2-\caP^{(1)}_+}.
\ee
We now prove the expression for $\caE^{(2)}_+$. In Figure \ref{subfig:E2_+} the trajectories contributing to $\caE^{(2)}_+$ are represented schematically. Suppose that the walker just entered the set $\mathbb{S}_+$, i.e. it is located at $x^{(1)}_{k-1}+1$. Then it can again move in and out of the green area an arbitrary number of times. As established before, the probability of moving in and out of the green area $n$ times is given by $(\caP^{(1)}_+/2)^n$. The expectation of the time spent in the green area after entering it for the $n$-th time is the probability that the walker survives moving in and out $n-1$ times, i.e. $(\caP^{(1)}_+/2)^{n-1}$, multiplied by the probability that the walker then jumps back in, i.e. ($1/2$), multiplied by the expectation of the time before leaving the green area plus one for the extra step entering, i.e. ($\caE^{(1)}_+ + 1$). The value of $\caE^{(2)}_+$ is the sum of these expectations of all these time intervals plus the probability that the walker leaves $\mathbb{S}_+$ before dying. This extra term is because of the extra time unit it costs to the walker to jump from $x_{k-1}+1$ to $x_{k-1}$,i.e. out of $\mathbb{S}_+$. We find

\be\label{eq:expressioncae}
\caE^{(2)}_+ = \frac{1}{2}(\caE^{(1)}_+ + 1) \sum_{n=0}^{\infty} \Big( \frac{1}{2} \caP^{(1)}_+\Big)^n + \caP^{(2)}_+ = \frac{\caE^{(1)}_+ +1}{2-\caP^{(1)}_+} + \frac{1}{2-\caP^{(1)}_+} = \frac{\caE^{(1)}_+ +2}{2-\caP^{(1)}_+}.
\ee
\epr

Using Lemma \ref{lemma:expression_suvival_curly} and Lemma \ref{Lemma:calculation_E_+^{(2)}}, we find the following expressions for $\E[\tau^{(1)}]$ and $\E[\tau^{(2)}]$ in terms of $\caE^{(1)}_\pm$ and $\caP^{(1)}_\pm$,

\be\label{eq:etau1}
\E[\tau^{(1)}] = \caE^{(1)}_0 +  \caP^{(1)}_0 \Big[ \caE^{(1)}_-  \frac{1}{1-\caP^{(1)}_+ \caP^{(1)}_-} + \caE^{(1)}_+ \frac{\caP^{(1)}_-}{1-\caP^{(1)}_+ \caP^{(1)}_-} \Big]
\ee

\begin{align}\label{eq:etau2}
\E[\tau^{(2)}] &= \caE^{(2)}_0 +  \caP^{(2)}_0 \Big[ \caE^{(2)}_-  \frac{1}{1-\caP^{(2)}_+ \caP^{(2)}_-} + \caE^{(2)}_+ \frac{\caP^{(2)}_-}{1-\caP^{(2)}_+ \caP^{(2)}_-} \Big]\\
&= \caE^{(1)}_0 +  \caP^{(1)}_0 \Big[ \caE^{(1)}_-  \frac{1}{1-\frac{1}{2-\caP^{(1)}_+} \caP^{(1)}_-} + \frac{\caE^{(1)}_+ + 2}{2- \caP^{(1)}_+} \frac{\caP^{(1)}_-}{1-\frac{1}{2-\caP^{(1)}_+}\caP^{(1)}_-} \Big]\nonumber\\
&= \caE^{(1)}_0 +  \caP^{(1)}_0 \Big[ \caE^{(1)}_-  \frac{2-\caP^{(1)}_+}{2- \caP^{1}_+ - \caP^{(1)}_-} + \caE^{(1)}_+ \frac{\caP^{(1)}_-}{2- \caP^{(1)}_+ - \caP^{(1)}_-} + \frac{2 \caP^{(1)}_-}{2-\caP^{(1)}_+ - \caP^{(1)}_-} \Big].\nonumber
\end{align}
We find that $\E[\tau^{(2)}] \geq \E[\tau^{(1)}]$ if and only if

\begin{align} \label{eq:condition_monotonicity}
    0 &\leq \frac{1}{\caP^{(1)}_0} \left(\E[\tau^{(2)}] - \E[\tau^{(1)}] \right)\\
    & = \caE^{(1)}_- \left[ \frac{2-\caP_+^{(1)}}{2-\caP_+^{(1)} - \caP_-^{(1)}} -  \frac{1}{1- \caP_+^{(1)}\caP_-^{(1)} }\right]\nonumber\\
    & \quad + \caE^{(1)}_+ \caP^{(1)}_-\left[ \frac{1}{2- \caP^{(1)}_+ - \caP^{(1)}_-} - \frac{1}{1- \caP^{(1)}_+ \caP^{(1)}_-}\right]\nonumber\\
    & \quad+ \frac{2 \caP^{(1)}_-}{2-\caP^{(1)}_+ - \caP^{(1)}_-}\nonumber
\end{align}

\subsection{Example of non-monotonicity} \label{subsec:counterexample}
We will now use this expression \eqref{eq:condition_monotonicity} to show that monotonicity does not hold for some class of trap landscapes. We choose $k = 2$ and $x^{(1)}_2 = x^{(1)}_1+1$. Below is depicted how such a landscape $\om^{(1)}$ and the associated landscape $\om^{(2)}$ can look.

\noindent
\underline{Landscape $\om^{(1)}$:}

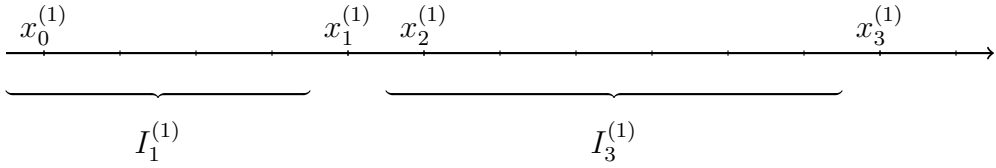
\begin{center}
\begin{tikzpicture}
\centering
\draw[thick,->] (-0.5,0) -- (12.5,0) ;
\foreach \x in {0,1,2,3,4,5,6,7,8,9,10,11,12}
   \draw (\x cm,1pt) -- (\x cm,-1pt);
   
\draw (0 cm,1pt) -- (0 cm,-1pt) node[anchor=south] {$x^{(1)}_{0}$};
\draw (4 cm,1pt) -- (4 cm,-1pt) node[anchor=south] {$x^{(1)}_{1}$};
\draw (5 cm,1pt) -- (5 cm,-1pt) node[anchor=south] {$x^{(1)}_{2}$};
\draw (11 cm,1pt) -- (11 cm,-1pt) node[anchor=south] {$x^{(1)}_{3}$};

\draw [thick, decorate, decoration = {calligraphic brace,mirror}] (-0.5,-0.5) --  (3.5,-0.5) node[pos=0.5,below = 7pt,black]{$I^{(1)}_{1}$};
\draw [thick, decorate, decoration = {calligraphic brace,mirror}] (4.5,-0.5) --  (10.5,-0.5) node[pos=0.5,below = 7pt,black]{$I^{(1)}_{3}$};
\end{tikzpicture}
\end{center}

\noindent
\underline{Landscape $\om^{(2)}$:}

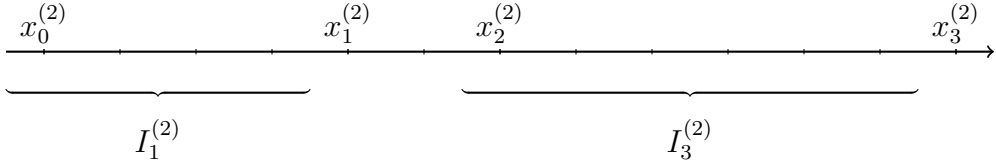
\begin{center}
\begin{tikzpicture}
\centering
\draw[thick,->] (-0.5,0) -- (12.5,0) ;
\foreach \x in {0,1,2,3,4,5,6,7,8,9,10,11,12}
   \draw (\x cm,1pt) -- (\x cm,-1pt);
   
\draw (0 cm,1pt) -- (0 cm,-1pt) node[anchor=south] {$x^{(2)}_{0}$};
\draw (4 cm,1pt) -- (4 cm,-1pt) node[anchor=south] {$x^{(2)}_{1}$};
\draw (6 cm,1pt) -- (6 cm,-1pt) node[anchor=south] {$x^{(2)}_{2}$};
\draw (12 cm,1pt) -- (12 cm,-1pt) node[anchor=south] {$x^{(2)}_{3}$};

\draw [thick, decorate, decoration = {calligraphic brace,mirror}] (-0.5,-0.5) --  (3.5,-0.5) node[pos=0.5,below = 7pt,black]{$I^{(2)}_{1}$};
\draw [thick, decorate, decoration = {calligraphic brace,mirror}] (5.5,-0.5) --  (11.5,-0.5) node[pos=0.5,below = 7pt,black]{$I^{(2)}_{3}$};
\end{tikzpicture}
\end{center}

We first notice that the following bounds hold for $\caP^{(1)}_+$, $\caP^{(1)}_-$ and $\caE^{(1)}_+$, $\caE^{(1)}_-$.

\be \label{eq:bound_P+_P-}
\frac{1}{3} \leq \caP^{(1)}_+ , \caP^{(1)}_- \leq \frac{2}{3}
\ee
and
\be \label{eq:bounds_E+_E-}
\caE^{(1)}_- \leq 2 |I^{(1)}_1| + 1 \qquad , \qquad \caE^{(1)}_+ \geq \frac{1}{3} (|I^{(1)}_3|-1) 
\ee
Equation \eqref{eq:bound_P+_P-} can be understood from the following observation. 
Suppose the walker 
is located at $x^{(1)}_1\in\mathbb{S}_-$. One possible way the walker can leave $\mathbb{S}_-$ before it gets trapped is to survive the trap at $x^{(1)}_1$ and next jump right out of $\mathbb{S}_-$, thus entering in $\mathbb{S}_+$. This happens with probability $1/3$, hence $1/3 \leq \caP^{(1)}_-=\Pr(l^{(1)}_{2m+1} < \infty| l^{(1)}_{2m} < \infty )$ for $m\geq 0$. One way to get trapped before moving out of $\mathbb{S}_-$ is to immediately get trapped on $x^{(1)}_1$ so that $l^{(1)}_{1} =\infty$. This happens with  probability $1/3$, hence $\caP^{(1)}_- \leq 1 - 1/3 = 2/3$. 
The bounds for $\caP^{(1)}_+$ can be constructed in the same way, because  by entering in $\mathbb{S}_+$ the walker is located on a trap, namely $x^{(1)}_2$, and since it entered in $\mathbb{S}_+$, we have $l^{(1)}_{2m+1} <\infty$ and we can bound $\caP^{(1)}_+=\Pr(l^{(1)}_{2m+2} < \infty| l^{(1)}_{2m+1} < \infty )$.

   The upper bound for  $\caE^{(1)}_-=\E[T^{(1)}_{2m+1} | l^{(1)}_{2m} < \infty]$ (for $m\geq 0)$  in \eqref{eq:bounds_E+_E-} can be understood by considering a random walker on $\N$ without the traps. That is, we calculate the expectation of the time it takes the random walker to leave the set $\mathbb{S}_-$ without there being any traps. This is an upper bound for $\caE^{(1)}_-$, because $\caE^{(1)}_-$ is the expectation of the minimum of the time to leave $\mathbb{S}_-$ and the time to get trapped. Using \eqref{eq:time_in_interval} we see that the expectation of the time it takes the random walker to leave the set $\mathbb{S}_-$ is $2|I^{(1)}_1|+1$.
Indeed, due to reflection at the origin this is the same as the expected time that the simple random walk on $\Z$ starting at $x_1$ hits $x_1+1$ before
$-x_1-1$. Then an application of the formula 
\eqref{eq:time_in_interval} gives $2x_1+1$ for this expected time.

For the lower bound on $\caE^{(1)}_+=\E[T^{(1)}_{2m} | l^{(1)}_{2m-1} < \infty ]$ (for $m \geq 0$)  we consider only the contribution of trajectories which survive $x^{(1)}_2$ after entering $\mathbb{S}_+$ and then jump right. Such a trajectory occurs with probability $1/3$. After that, we bound $\caE^{(1)}_+$ by the expectation of the time at which the walker encounters the next trap (either $x^{(1)}_2$ or $x^{(1)}_3$), which is $|I_3|-1$ by \eqref{eq:time_in_interval}.

One can, e.g. numerically, check that there exist positive constants $b,B \in \R^+$ and $c \in \R^+$ such that

\be\label{eq:num1}
B \geq \frac{2-\caP_+^{(1)}}{2-\caP_+^{(1)} - \caP_-^{(1)}} -  \frac{1}{1- \caP_+^{(1)}\caP_-^{(1)} } > 0,
\ee
\be\label{eq:num2}
   \frac{1}{2- \caP^{(1)}_+ - \caP^{(1)}_-} - \frac{1}{1- \caP^{(1)}_+  \caP^{(1)}_-} \leq -b, < 0 
\ee
\be\label{eq:num3}
0 < \frac{2 \caP^{(1)}_-}{2-\caP^{(1)}_+ - \caP^{(1)}_-} < c,
\ee
for all values of $\caP^{(1)}_-, \caP^{(1)}_+$
in $[1/2,2/3]$. The right hand side of \eqref{eq:condition_monotonicity} can be bounded from above by
\be\label{eq:num4}
\left(2|I^{(1)}_1|+1 \right) B - \frac{1}{3}\left(|I^{(1)}_3|-1\right)  b + c.
\ee
Now we  see that \eqref{eq:condition_monotonicity} doesn't generally hold. Indeed, we can choose $|I^{(1)}_1|$ and $|I^{(1)}_3|$ such that the inequality in \eqref{eq:condition_monotonicity} is violated.

\subsection{Proof of Theorem \ref{thm:finiteness_tail}} \label{subsec:proof_finiteness_tail}
We prove Theorem \ref{thm:finiteness_tail} using the tools we developed for showing that monotonicity doesn't generally hold. That is, we show that for two trap landscapes $\om^{(a)}$ and $\om^{(b)}$ which are similar in the tail we have

\[
\E[\tau^{(a)}] < \infty \qquad \text{if and only if} \qquad \E[\tau^{(b)}] < \infty.
\]
Here, $\tau^{(a)}$ and $\tau^{(b)}$ denote the survival times associated to landscape $\om^{(a)}$ and landscape $\om^{(b)}$ respectively. As for the survival time, we will write superscript $(a)$ to indicate quantities that correspond to $\om^{(a)}$ and superscript $(b)$ to indicate quantities that correspond to $\om^{(b)}$. Recall that the landscapes $\om^{(a)}$ and $\om^{(b)}$ are similar in the tail if there exist $p_a,p_b \in \N$ such that

\[
\om^{(a)}(p_a + x) = \om^{(b)}(p_b + x)
\]
for all $x \in \N$. We define $k^{(a)} = \inf\{k \in \N: x^{(a)}_{k-1} \geq p_a\}$, i.e. $k^{(a)}-1$ is the index of the first trap to the right of (or on) site $p_a$ in $\om^{(a)}$. Similarly we define $k^{(b)} = \inf \{k \in \N: x^{(b)}_{k-1} \geq p_b\}$, such that $k^{(b)}-1$  is the index of the trap to the right of (or on) site $p_b$ in $\om^{(b)}$. We construct the quantities $\caP^{(1)}_0, \caP^{(1)}_-, \caP^{(1)}_+$ and $\caE^{(1)}_0, \caE^{(1)}_-, \caE^{(1)}_+$ for $\om^{(a)}$ and $k^{(a)}$ and denote these $\caP^{(a)}_0, \caP^{(a)}_-, \caP^{(a)}_+$ and $\caE^{(a)}_0, \caE^{(a)}_-, \caE^{(a)}_+$. By Lemma \ref{lemma:expression_suvival_curly} we have
\be \label{eq:E_tau_a}
\E[\tau^{(a)}] = \caE^{(a)}_0 +  \caP^{(a)}_0 \Big[ \caE^{(a)}_-  \frac{1}{1-\caP^{(a)}_+ \caP^{(a)}_-} + \caE^{(a)}_+ \frac{\caP^{(a)}_-}{1-\caP^{(a)}_+ \caP^{(a)}_-} \Big].
\ee
From \eqref{eq:E_tau_a} we can see that $\E[\tau^{(a)}]$ is finite if and only of $\caE^{(a)}_+$ is finite. Indeed, all other variables in $\eqref{eq:E_tau_a}$ are finite and $\caP_0^{(a)},\caP_-^{(a)}>0$. Similarly, we construct the quantities $\caP^{(1)}_0, \caP^{(1)}_-, \caP^{(1)}_+$ and $\caE^{(1)}_0, \caE^{(1)}_-, \caE^{(1)}_+$ for $\om^{(b)}$ and $k^{(b)}$ and denote these $\caP^{(b)}_0, \caP^{(b)}_-, \caP^{(b)}_+$ and $\caE^{(b)}_0, \caE^{(b)}_-, \caE^{(b)}_+$. As for $\E[\tau^{(a)}]$, we have that $\E[\tau^{(b)}]$ is finite if and only if $\caE^{(b)}_+$ is finite. 

Notice that $\caE^{(a)}_{+}$ only depends on the distances between the traps
\be \label{eq:traps_om_a}
x^{(a)}_{k^{(a)}-1}, x^{(a)}_{k^{(a)}}, x^{(a)}_{k^{(a)}+1}, \dots
\ee 
and $\caE^{(b)}_{+}$ only depends on the distances between the traps 
\be\label{eq:traps_om_b}
x^{(b)}_{k^{(b)}-1}, x^{(b)}_{k^{(b)}}, x^{(b)}_{k^{(b)}+1}, \dots.
\ee
Because $\om^{(a)}$ and $\om^{(b)}$ are similar in the tail and $k^{(a)}, k^{(b)}$ are defined such that both
\be\label{suchthatboth}
x_{k^{(a)} - 1} \geq p_a \qquad \text{and} \qquad x_{k^{(b)} - 1} \geq p_b,
\ee
we have that these distances between the traps in \eqref{eq:traps_om_a} and \eqref{eq:traps_om_b} are the same. We conclude that $\caE^{(a)}_+$ and $\caE^{(b)}_+$ are the same. As a consequence both $\E[\tau^{(a)}]$ and $\E[\tau^{(b)}]$ are finite if and only if $\caE^{(a)}_+ (=\caE^{(b)}_+)$ is finite.

\section{Appendix}\label{sec:appendix}
We show that under the restriction
$(c|I_1|)^{2^n-1} |I_1|   \in \N$
the constant $c$ is of the form
\be
c = \frac{\gamma}{|I_1|}
\ee
for some $\gamma\in \N$. For $n=1$ we have

\be \label{eq:appendix_1}
(c|I_1|)^{2^n-1} |I_1| = c|I_1|^2 \in \N.
\ee
Because $|I_1| \in \N$, this implies that $c |I_1|$ is rational. There exist $a,b \in \N$ which are coprime such that
\be
c |I_1| = \frac{a}{b}.
\ee
We will show that $c|I_1|$ is in fact an integer, i.e. $b=1$. Equation \eqref{eq:appendix_1} becomes
\be \label{eq:appendix_2}
(c|I_1|)^{2^n-1} |I_1| = \left( \frac{a}{b}\right)^{2^n-1} |I_1| \in \N.
\ee
Consider the prime decompositions of $a$, $b$ and $|I_1|$,
\begin{align}
    a &= a_1 \dots a_k\\
    b &= b_1 \dots b_m \nonumber \\
    |I_1| &= i_1 \dots i_l.\nonumber
\end{align}
The right hand side of \eqref{eq:appendix_2} becomes
\be
(a_1 \dots a_k)^{2^n-1} \cdot \frac{i_1 \dots i_l}{(b_1 \dots b_m)^{2^n-1}} \in \N.
\ee
For this statement to hold, the factors $b_1^{2^n-1}, \dots b_m^{2^n-1}$ in the denominator must all cancel with factors in the numerator $(a_1 \dots a_k)^{2^n-1} \cdot i_1 \dots i_l$. Moreover, because $a$ and $b$ are coprime, the factors $b_1^{2^n-1}, \dots b_m^{2^n-1}$ must cancel with $i_1,...,i_l$. Since $n$ can be arbitrarily large, this is only possible if $b_1 = \dots = b_m = 1$. This shows that $b=1$ and hence $c|I_1| = a/b = a$ is an integer.\\ \\

\noindent
\textbf{Acknowledgements.} We thank Adrien Rezzouk for useful discussions, in particular for pointing us the non-monotonicity aspect explained in Example \ref{ex:nonintuitive}. F.R. and B.v.T.
thank Universit\'e Paris Cit\'e and Laboratoire MAP5 for financial support and hospitality. B.v.T. is supported by the Peter Paul Peterich Foundation via TU Delft University Fund.

\bibliography{biblio}
\bibliographystyle{unsrt}

\end{document}